%% file: GSIS_HOLO.tex
\documentclass[onefignum,onetabnum]{siamart171218}

\usepackage{braket,amsfonts}
\usepackage{array}
\usepackage[caption=false]{subfig}
\usepackage{pgfplots,bm,multirow}

\newsiamthm{claim}{Claim}
\newsiamremark{remark}{Remark}
\newsiamremark{hypothesis}{Hypothesis}
\crefname{hypothesis}{Hypothesis}{Hypotheses}

\usepackage{algorithmic}

\usepackage{graphicx,epstopdf}

\usepackage{amsopn}

\usepackage{xspace}
\usepackage{bold-extra}
\usepackage[most]{tcolorbox}

\colorlet{texcscolor}{blue!50!black}
\colorlet{texemcolor}{red!70!black}
\colorlet{texpreamble}{red!70!black}
\colorlet{codebackground}{black!25!white!25}

\lstdefinestyle{siamlatex}{%
	style=tcblatex,
	texcsstyle=*\color{texcscolor},
	texcsstyle=[2]\color{texemcolor},
	keywordstyle=[2]\color{texemcolor},
	moretexcs={cref,Cref,maketitle,mathcal,text,headers,email,url},
}

\tcbset{%
	colframe=black!75!white!75,
	coltitle=white,
	colback=codebackground, 
	colbacklower=white, 
	fonttitle=\bfseries,
	arc=0pt,outer arc=0pt,
	top=1pt,bottom=1pt,left=1mm,right=1mm,middle=1mm,boxsep=1mm,
	leftrule=0.3mm,rightrule=0.3mm,toprule=0.3mm,bottomrule=0.3mm,
	listing options={style=siamlatex}
}

\newtcblisting[use counter=example]{example}[2][]{%
	title={Example~\thetcbcounter: #2},#1}

\newtcbinputlisting[use counter=example]{\examplefile}[3][]{%
	title={Example~\thetcbcounter: #2},listing file={#3},#1}

\DeclareTotalTCBox{\code}{ v O{} }
{ 
	fontupper=\ttfamily\color{black},
	nobeforeafter,
	tcbox raise base,
	colback=codebackground,colframe=white,
	top=0pt,bottom=0pt,left=0mm,right=0mm,
	leftrule=0pt,rightrule=0pt,toprule=0mm,bottomrule=0mm,
	boxsep=0.5mm,
	#2}{#1}

\patchcmd\newpage{\vfil}{}{}{}
\begin{tcbverbatimwrite}{tmp_\jobname_header.tex}
	\title{General synthetic iterative scheme for unsteady rarefied gas flows
		 \thanks{Submitted to the editors DATE.
	}}
	
	\author{
		 Lei Wu\thanks{Department of Mechanics and Aerospace Engineering, Southern University of Science and Technology, Shenzhen 518055, China (\email{wul@sustech.edu.cn}) }
	}

	\headers{Unsteady GSIS: fast convergence and asymptotic  preserving}{Lei Wu}
\end{tcbverbatimwrite}
\input{tmp_\jobname_header.tex}

\ifpdf
\hypersetup{ pdftitle={Guide to Using SIAM'S \LaTeX\ Style} }
\fi

\definecolor{mygreen}{rgb}{0.0,0.55,0.55}

\begin{document}
	\maketitle

	\begin{tcbverbatimwrite}{tmp_\jobname_abstract.tex}
		\begin{abstract}
	In rarefied gas flows, the spatial grid size could vary by several orders of magnitude in  a single flow configuration (e.g., inside the Knudsen layer it is at the order of mean free path of gas molecules, while in the bulk region it is at a much larger hydrodynamic scale). Therefore, efficient implicit numerical method is urgently needed for time-dependent problems. 
	However, the integro-differential nature of gas kinetic equations poses a grand challenge, as the gain part of the collision operator is non-invertible. Hence an iterative solver is required in each time step, which usually takes a lot of iterations in the (near) continuum flow regime where the Knudsen number is small; worse still, the solution does not asymptotically preserve the fluid dynamic limit when the spatial cell size is not refined enough.
Inspired by our general synthetic iteration scheme for steady-state solution of the Boltzmann equation, we propose two numerical schemes to push the multiscale simulation of unsteady rarefied gas flows to a new boundary, that is, the numerical solution not only converges within dozens of iterations in each time step, but also asymptotic preserves the Navier-Stokes-Fourier limit at coarse spatial grid, even when the time step is large (e.g., in the extreme slow decay of two-dimensional Taylor vortex, the time step is at the order of vortex decay time). The properties of fast convergence and asymptotic preserving of the proposed schemes are not only rigorously proven by the Fourier stability analysis, but also demonstrated by solid numerical examples.	
		\end{abstract}
		
		
	\end{tcbverbatimwrite}
	\input{tmp_\jobname_abstract.tex}

\section{Introduction}

Rarefied gas flows have attracted significant research interest in the past decades due their wide range of  engineering applications. These flows are characterized by the Knudsen number $\text{Kn}$, which is defined as the ratio of  mean free path $\lambda$ (or mean collision time $t_c$) of gas molecules to the characteristic flow length $L$ (or time/period $T$). Only when the Knudsen number is small can the rarefied gas dynamics be well described by macroscopic equations in the bulk flow region~\cite{Sone2002Book}, such as the Euler, Navier-Stokes-Fourier (NSF), Burnett~\cite{CE}, and Grad 13 moments equations~\cite{Grad1949}; see the short review on the performance of macroscopic equations~\cite{Wu2020AIA}. For general values of Knudsen number, however, the Boltzmann equation or simplified gas kinetic equations should be used.

Since the velocity distribution function is defined in a six-dimensional phase space, the computational cost of memory and time in the solving of gas kinetic equations is huge. Thus, many numerical methods are proposed to solve the kinetic equations under a numerical scale larger than the kinetic one~\cite{Luc2000JCP,Filbet2010JCP,Filbet2011,Dimarco2017Siam,Ju2017JSC,UGKS2010JCP,guo2013discrete}, that is, the spatial grid size $\Delta{x}\gg\lambda$, and/or the time step $\Delta{t}\gg\tau_c$. Some schemes asymptotically preserve the Euler limit, as they become a consistent discretization of the Euler equations when $\text{Kn}\rightarrow0$~\cite{Filbet2010JCP,Filbet2011}. 
Nevertheless, from a practical point of view, the Euler equations cannot be applied for some flows, even when the Knudsen number is very small. For instances, in the Poiseuille flow~\cite{WANG201833} and Rayleigh-Brillouin scattering~\cite{Su2020SIAM}, the flow velocity and density amplitude scales as $1/\text{Kn}$. If the Euler equations are used, they become divergent, which is not physical.
Therefore, some numerical schemes are designed to asymptotically preserve the NSF limit when $\Delta{t}\gg\tau_c$~\cite{Dimarco2017Siam,Ju2017JSC}, under the assumption that the spatial derivatives are handled exactly. Recently, it is found that the NSF limit can be captured by the (discrete) unified gas-kinetic scheme,  when both the time step and spatial cell size are much larger than the corresponding kinetic scales~\cite{UGKS2010JCP,guo2013discrete,Guo2019UP_arXiv,Zhu2019JCP}, e.g., $\Delta{x}\sim\sqrt{\text{Kn}}L\gg\lambda$. 
It should be noted that, however, these large numerical scales are used in bulk flow region only, without resolving the Knudsen layer which occupies a spatial region within a few mean free path away from the solid walls.

In reality, rarefied gas flows are intrinsically multiscale, say, in the two-dimensional thermal edge flow in the (near) continuum flow regime where the Knudsen number is small~\cite{Su2020SIAM}, the spatial grid size varies by several orders of magnitude: inside the Knudsen layer $\Delta{x}\sim\lambda\sim0.001$, while in the bulk region it is at a much larger hydrodynamic scale: $\Delta{x}\sim{}L\sim1$. Such a disparate grid cell size is necessary, as the fine grid in the Knudsen layer helps to capture the ghost effect that arises from the rarefaction effects inside the Knudsen layer and determines the velocity field in the whole computational domain~\cite{Yoshio2000}, while the coarse grid help to save the computational cost; our numerical tests show that, if the Knudsen layer is under-resolved, the vortexes  rotate in wrong directions. Therefore, for time-dependent problems, it is highly desired to use the implicit numerical method, otherwise the time step in explicit numeral methods will be restricted by the Courant–Friedrichs–Lewy condition, namely, $\Delta{t}\sim\tau_c$, which renders any practical numerical simulation impossible when the Knudsen number is small. 


However, a grand challenge arises for implicit numerical method, since the integro-differential nature of gas kinetic equations makes the gain part of the collision operator non-invertible. Therefore, an iterative solver is required in each time step. A simple way is to use the conventional iterative scheme (CIS), where the time derivative, streaming operator, and loss part of the collision operator are evaluated at the current iteration step, while the gain part of the collision operator is obtained from the previous iteration step. The CIS is efficient when the Knudsen number is large, but it takes a lot of iterations in the (near) continuum flow regime; worse still, the CIS does not asymptotically preserve the NSF limit, so a huge number of spatial cells should be used to capture the gas dynamics~\cite{WANG201833}. To improve the efficiency of implicit iteration, a fast convergence and asymptotic NSF preserving  scheme is urgently needed.

Thanks to the pioneering work of Larsen in neutron transport~\cite{Larsen1983NSE}, the acceleration is possible if the kinetic equation and its moment equations are coupled: the kinetic equation provides high-order moments to moment equations (diffusion equations in the content of neutron transport), while the moment equations provide macroscopic quantities appearing in the gain part of the collision operator. 
The essential idea was extended to gas kinetic system, such as the high-order/low-order (HOLO) method~\cite{Taitano2014} for the Bhatnagar-Gross-Krook (BGK) model~\cite{Bhatnagar1954} and the general synthetic iterative scheme (GSIS) for the Boltzmann equation~\cite{SuArXiv2019}. In the Sod shock tube problem, it was found that in each time step HOLO finds the converged solution within dozens of iterations~\cite{Taitano2014}. Albeit this promising property, the stability and asymptotic NSF preserving of HOLO have not been rigorously analyzed. On the other hand, while the fast convergence and asymptotic NSF preserving of GSIS has been proven~\cite{Su2020SIAM,Su2021CMAME,Zhu2021JCP}, so far it has been limited to steady-state problems.


In this paper we extend the GSIS to time-dependent problems, retaining its unique properties of fast convergence and asymptotic NSF preserving.
In GSIS, the kinetic equation and macroscopic synthetic equations can be solved by different numerical methods with different order of accuracy. This brings tremendous numerical convenience because the kinetic equation, which requires discretization in the high-dimensional phase space, are usually time-consuming and hence should be handled by as simple algorithm as possible. The macroscopic synthetic equation, on the contrary, are well studied in computational fluid dynamics and can be handled by sophisticated high-order numerical methods.   
From the practical application point of view, we hope the developed kinetic schemes could use as large time step and spatial cell size as possible. That is, $\Delta{t}$ and $\Delta{x}$ can be chosen as the maximum ones used in the NSF equations to capture continuum flow dynamics accurately.

In section~\ref{kinetic_equations} the gas kinetic equation and its asymptotic behavior at small Knudsen number is introduced. In section~\ref{section_CIS} the CIS to solve the time-dependent gas kinetic equation is introduced and the convergence rate of iterations is calculated by the Fourier stability analysis, at the whole range of Knudsen number and time step. In sections~\ref{section_GSIS2} and~\ref{section_GSIS1} two general synthetic iterative schemes are proposed and their convergence rates are calculated. That of HOLO is also calculated and compared. In section~\ref{section_AP} the conditions for asymptotic NSF preserving are analyzed. Several challenging numerical examples are used to assess the accuracy and efficiency of the proposed GSIS for time-dependent problems in section~\ref{section_numerical}. 
Finally, a summary is given in section~\ref{section_summary}.

\section{Kinetic equations and moment equations}\label{kinetic_equations}

We take the linearized gas kinetic models to illustrate how the GSIS can be extended to unsteady problems; the extension to nonlinear Boltzmann equation is straightforward~\cite{SuArXiv2019,Zhu2021JCP}. In the absence of external acceleration, the kinetic model reads
\begin{equation}\label{bgkfd}
\frac{\partial h}{\partial t}
+\bm{v} \cdot  \frac{\partial h}{\partial \bm{x}} =\mathcal{L}_s(h)\equiv\mathcal{L}^{+}- \delta_{rp}  h. 
\end{equation}
Here, $h(t,\bm{x},\bm{v})$ is the perturbation velocity distribution function that depends on the time $t\in\mathbb{R}^+$, molecular velocity $\bm{v}=(v_1,v_2,v_3)\in\mathbb{R}^3$, and spatial coordinates $\bm{x}=(x_1,x_2,x_3)\in\mathbb{R}^3$. 
The rarefaction parameter $\delta_{rp}$ (its inverse is known as the Knudsen number $\text{Kn}$) is a constant in linearized problems. When $\delta_{rp}$ is large, the kinetic model becomes stiff, which requires special care to make the solution converge fast and asymptotically preserve the NSF limit even with coarse spatial grid and large temporal step. The loss part of the collision operator is $\delta_{rp}h$, while the gain part is~\cite{Shakhov1968} 
\begin{equation}\label{LBE_Shakhov}
\mathcal{L}^+=\delta_{rp}\left[\varrho+2\bm{u}\cdot\bm{v}+\tau\left(v^2-\frac{3}{2}\right)+\frac{4(1-\text{Pr})}{5}\bm{q}\cdot{\bm{v}}\left(v^2-\frac{5}{2}\right)\right]f_{eq},
\end{equation} 
where the global equilibrium velocity distribution function is
\begin{equation}
f_{eq}=\frac{1}{\pi^{3/2}}\exp(-v^2),
\end{equation}
and
the macroscopic quantities, such as the perturbation density $\rho$, flow velocity $\bm{v}$, perturbation temperature $\tau$, stress $\sigma_{ij}$, and heat flux $\bm{q}$ are defined as the velocity moments of the velocity distribution function: 
\begin{equation}\label{MP}
\begin{aligned}[c]
\varrho=\int{h}d\bm{v}, \quad 
\bm{u}=\int{\bm{v}h}d\bm{v}, \quad 
\tau=\frac{2}{3}\int\left({v^2}-\frac{3}{2}\right)hd\bm{v}, \\
\sigma_{ij}=2\int{v_{\langle{i}}v_{j\rangle}h}d\bm{v}
\equiv{}2\int\left(v_iv_j-\frac{v^2}{3}\delta_{ij}\right){h}d\bm{v}, \quad
\bm{q}=\int\left({v^2}-\frac{5}{2}\right)\bm{v}hd\bm{v}.
\end{aligned}
\end{equation}
The kinetic model reduces to the linearized BGK~\cite{Bhatnagar1954} and Shakhov~\cite{Shakhov_S} models when the Prandtl number is $\text{Pr}=1$ and $2/3$, respectively.  For real monatomic gas, the Prandtl number is approximately 2/3. Note that in the BGK model, the heat flux does not appear in the collision operator~\eqref{LBE_Shakhov}.

Note that in the kinetic model~\eqref{bgkfd} the spatial coordinates and molecular velocity have been normalized by the characteristic flow length $L$ and most probable speed $v_m=\sqrt{2RT_0}$, where $R$ is the gas constant and $T_0$ is the reference temperature. Also, the time has been normalized by $L/v_m$.

On multiply Eq.~\eqref{bgkfd} by 1, 2$\bm{v}$, and $v^2-\frac{3}{2}$, respectively, and integrate the resultant equations with respect to the molecular velocity $\bm{v}$, we obtain the following equations for the evolution of density, velocity, and temperature: 
\begin{equation}\label{eq123}
\begin{aligned}
\frac{\partial {\varrho}}{\partial{t}}+\frac{\partial {u_i}}{\partial{x_i}}=0, \\
2\frac{\partial {u_i}}{\partial{t}}+\frac{\partial {\varrho}}{\partial{x_i}}+\frac{\partial {\tau}}{\partial{x_i}}+\frac{\partial {{\sigma_{ij}}}}{\partial{x_j}}=0, \\
\frac{3}{2}\frac{\partial {\tau}}{\partial{t}}+\frac{\partial {{q_j}}}{\partial{x_j}}+\frac{\partial {u_j}}{\partial{x_j}}=0,
\end{aligned}
\end{equation}
which are not closed since the expressions for stress $\sigma_{ij}$ and heat flux $\bm{q}$ are not known. To close these moment systems, one either uses the Chapman-Enskog expansion~\cite{CE} or the Grad moment method~\cite{Grad1949}. However, the resultant equations are usually limited to finite value of Knudsen number. For instance, to the first order of Knudsen number, the constitutive relation is exactly the Newton's law for shear stress and Fourier's law for heat conduction:
\begin{equation}\label{NSF_relations_2}
\begin{aligned}[c]
\sigma_{ij}& =-2\delta_{rp}^{-1}\frac{\partial u_{<i}}{\partial {x_{j>}}}\equiv
-\delta_{rp}^{-1}\left(\frac{\partial u_{i}}{\partial x_{j}}+\frac{\partial u_{j}}{\partial x_{i}}-\frac{2}{3}\frac{\partial u_{k}}{\partial x_{k}}\delta_{ij}\right), \
q_i =-\frac{5}{4\text{Pr}}\delta_{rp}^{-1}\frac{\partial \tau}{\partial x_i}.
\end{aligned}
\end{equation} 
To describe the rarefied gas dynamics over the entire region of Knudsen numbers, it is necessary to solve the gas kinetic equation numerically.

\section{Conventional iterative scheme}\label{section_CIS}

A direct method to solve the kinetic equation, which is a complicated integro-differential equation, is the use of  conventional iteration scheme. In this paper, we consider the following typical temporal discretization of Eq.~\eqref{bgkfd} for unsteady problems: 
\begin{equation}\label{Shakhov_time_dependent0}
\frac{h_{n+1}-h_n}{\Delta{t}}+
\frac{\bm{v}}{2}\cdot
	\left( \frac{\partial{h_{n+1}}}{\partial\bm{x}}
	+ \frac{\partial{h_{n}}}{\partial\bm{x}}\right) =r\mathcal{L}_{n+1}+(1-r)\mathcal{L}_{n}, 
\end{equation}
where the quantities with the subscript $n$ are evaluated at the time $t_n$, $\Delta{t}=t_{n+1}-t_{n}$ is the time step, and the parameter $r$ varies between 0.5 and 1~\cite{Taitano2014,Zhu2019JCP}. When $r=1/2$, the scheme is second-order accuracy in time, while when $r=1$ it is a backward Euler scheme with first-order temporal accuracy.

Since $\mathcal{L}_{n+1}$ is a function of $h_{n+1}$, Eq.~\eqref{Shakhov_time_dependent0} must be solved iteratively. In CIS, given the value of velocity distribution function $h_{n+1}^{k}$ at the $k$-th iteration step (this is often called inner iteration in time-dependent implicit schemes~\cite{Zhu2019JCP}), its value at the next iteration step is calculated by:
\begin{equation}\label{Shakhov_time_dependent}
\frac{h^{k+1}_{n+1}-h_n}{\Delta{t}}+
\frac{\bm{v} }{2}\cdot
\left( \frac{\partial{h^{k+1}_{n+1}}}{\partial\bm{x}}
+ \frac{\partial{h_{n}}}{\partial\bm{x}}\right) =r\left({\mathcal{L}_{n+1}^{+,k}- \delta_{rp}  h_{n+1}^{k+1}}\right)
+(1-r){ \mathcal{L}_{n} },
\end{equation}
and this process repeats until the relative difference in macroscopic quantities between two consecutive inner iterations are less than a fixed value.

We adopt the Fourier stability analysis to investigate the efficiency of this inner iteration, that is, to see how fast the error decays when $k$ increases. We define the error functions between velocity distribution functions at two consecutive inner iterations as
\begin{equation}\label{Diff_Y}
Y^{k+1}(\bm{x},\bm{v})=h_{n+1}^{k+1}(\bm{x},\bm{v})-h_{n+1}^{k}(\bm{x},\bm{v}), 
\end{equation}
and the error functions for macroscopic quantities $M=[\varrho,\bm{u}, \tau,\bm{q}]$ between two consecutive inner iteration steps as 
\begin{equation}\label{Macro_difference}
\begin{aligned}
\Phi^{k+1}(\bm{x})\equiv&
\left[\Phi^{k+1}_\varrho,\Phi^{k+1}_{\bm{u}}, \Phi^{k+1}_\tau,\Phi^{k+1}_{\bm{q}}\right]\\
=&M_{n+1}^{k+1}(\bm{x}\,)-M_{n+1}^{k}(\bm{x}\,)=\int{Y^{k+1}(\bm{x},\bm{v}\,)\phi(\bm{v}\,)}d\bm{v},
\end{aligned}
\end{equation}
where
\begin{equation}
\begin{aligned}[c]
\phi(\bm{v}\,)=&\left[1,v_1,v_2,v_3,\frac{2}{3}v^2-1,v_1\left(v^2-\frac{5}{2}\right),v_2\left(v^2-\frac{5}{2}\right),
v_3\left(v^2-\frac{5}{2}\right)\right].
\end{aligned}
\end{equation}

From Eq.~\eqref{Shakhov_time_dependent}, it can be easily found that $Y^{k+1}(\bm{x},\bm{v})$ satisfies
\begin{equation}\label{YYY}
\begin{aligned}[c]
\left(1+\frac{1}{r\Delta{t}\delta_{rp}}+\frac{1}{2r\delta_{rp}} \bm{v} \cdot \bm \nabla\right)Y^{k+1}
=\\
	 f_{eq}
	 \left[\Phi^{k}_\varrho+2\Phi^{k}_{\bm{u}}\cdot\bm{v}
	 +\Phi^{k}_\tau\left(v^2-\frac{3}{2}\right)
	 +\frac{4(1-\text{Pr})}{5}\Phi^{k}_{\bm{q}}\cdot{\bm{v}}\left(v^2-\frac{5}{2}\right)\right],
\end{aligned}	 
\end{equation}
because in the $(n+1)$-th time step, all variables in the $n$-th time step are fixed and hence eliminated.


To determine the error decay rate $e$ we perform the Fourier stability analysis by seeking the eigenfunctions $\bar{Y}(\bm{v}\,)$ and $\alpha=[\alpha_\varrho,\bm\alpha_{u},  \alpha_{\tau},\bm\alpha_{q}]$ of the following forms:
\begin{equation}\label{ansatz}
\begin{aligned}[c]
Y^{k+1}(\bm{x},\bm{v}\,)=e^{k}\bar{Y}(\bm{v}\,)\exp(i\bm{\theta}\cdot{\bm{x}}\,),\quad
\Phi^{k+1}(\bm{x}\,)=e^{k+1}\alpha\exp(i\bm{\theta}\cdot{\bm{x}}\,),
\end{aligned}
\end{equation}
where $\bm{\theta}=(\theta_1,\theta_2,\theta_3)$ is the wavevector of perturbance and $i$ is the imaginary unit. The iteration is unstable when the error decay rate is larger than unity, while slow (fast) convergence occurs when the error decay rate $|e|$ approaches one (zero). Note that the two exponents in the right-hand-side of Eq.~\eqref{ansatz} are different, due to the fact that in CIS we first need macroscopic quantities to start the iteration. 

\begin{figure}[t]
	\centering
	\includegraphics[scale=0.45]{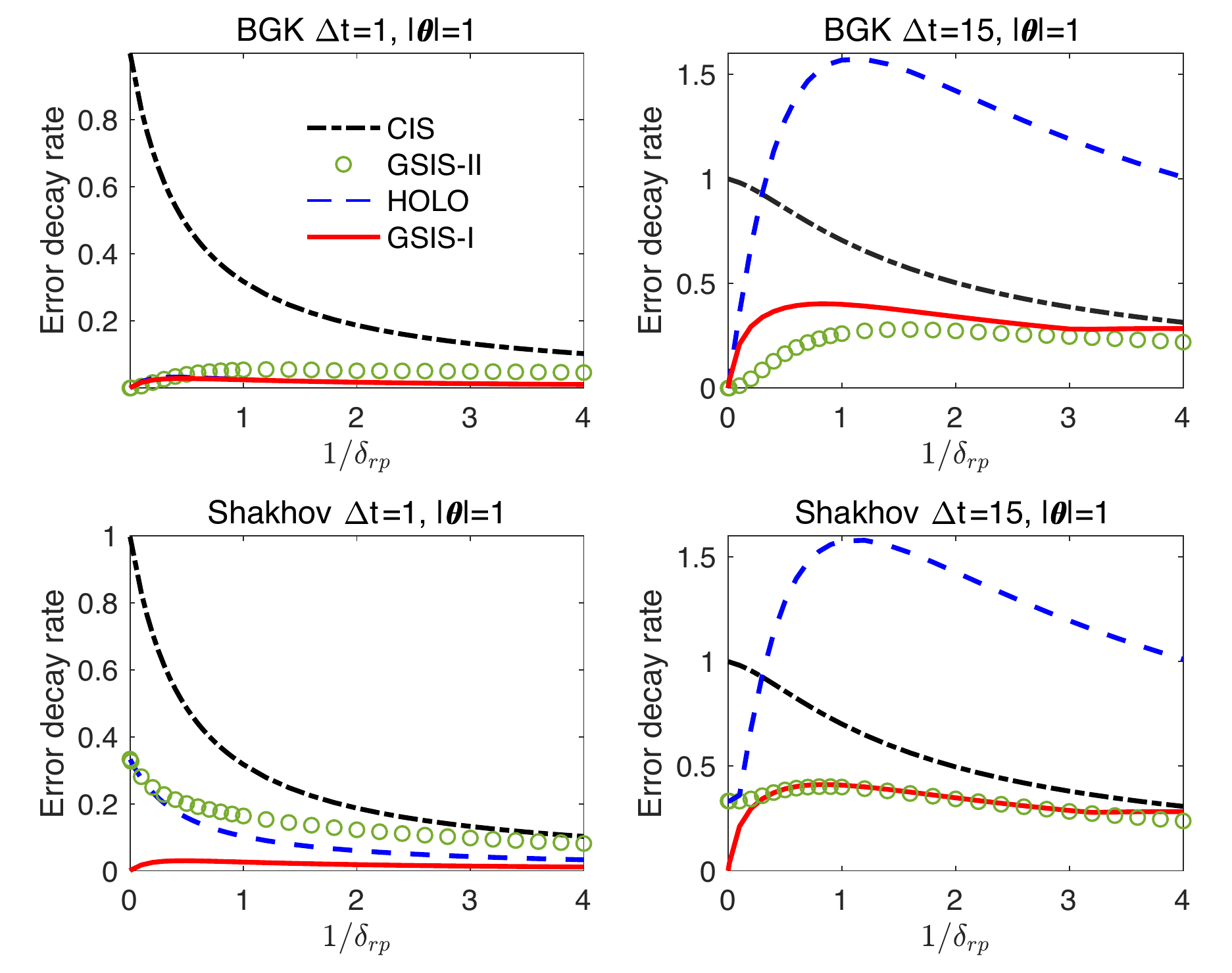}\\
	\includegraphics[scale=0.45]{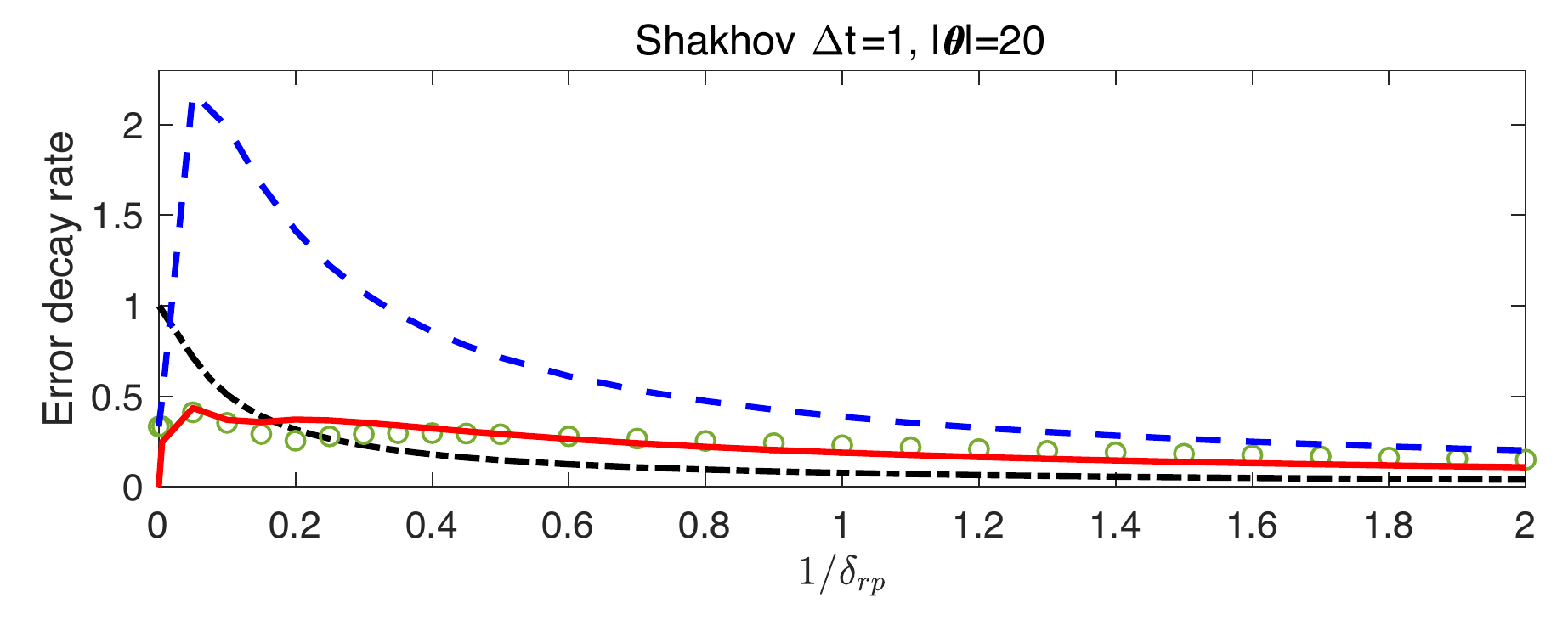}
	\caption{
		The error decay rate as a function of the  Knudsen number in CIS, GSIS, and HOLO~\cite{Taitano2014}, when $r=1/2$. Note that the iteration is unstable when the error decay rate is larger than one. 
	}
	\label{fig:GSIS12_time_dependent}
\end{figure}

Obviously, from Eqs.~\eqref{Macro_difference} and~\eqref{ansatz} we have 
\begin{equation}\label{relation}
e\alpha=\int \bar{Y}(\bm{v}\,)\phi(\bm{v}\,)d\bm{v},
\end{equation}
and from Eqs.~\eqref{YYY} and \eqref{ansatz}, we have
\begin{equation}\label{y0_solution_CIS_time_dependent}
\begin{aligned}[c]
\bar{Y}(\bm{v})=&\left[ \alpha_\varrho+2\bm\alpha_{u}\cdot\bm{v}+\alpha_{\tau}\left(v^2-\frac{3}{2}\right)+\frac{4(1-\text{Pr})}{5}\bm\alpha_{q}\cdot\bm{v}\left({v}^2-\frac{5}{2}\right) \right] y_0(\bm{v}),
\end{aligned}
\end{equation}
where
\begin{equation}
y_0(\bm{v})=\frac{{f_\text{eq}}}{ 1+(r\Delta{t}\delta_{rp})^{-1}+i(2r\delta_{rp})^{-1}\bm{\theta}\cdot\bm{v} } .
\end{equation}

On multiplying Eq.~\eqref{y0_solution_CIS_time_dependent} with $\phi(\bm{v})$ and integrating the resultant equations with respect to the molecular velocity $\bm{v}$, we obtain eight linear algebraic equations for eight unknown elements in $\alpha$ with the help of Eq.~\eqref{relation}. These algebraic equations can be written in the matrix form as
$
C_8\alpha^\top=e\alpha^\top$, 
where the superscript $\top$ is the transpose operator, and
$8\times8$ matrix is
\begin{equation}
C_8=\int \left[1,2\bm{v},v^2-\frac{3}{2},\frac{4(1-\text{Pr})}{5}\bm{v}\left(v^2-\frac{5}{2}\right)
\right]^\top \phi(\bm{v})y_0(\bm{v})d\bm{v}.
\end{equation}

The error decay rate can be obtained by numerically computing the eigenvalues of the matrix $C_8$ and taking the maximum absolute value of $e$;  results as a function of the Knudsen number, for both the BGK and Shakhov models, are shown in Fig.~\ref{fig:GSIS12_time_dependent}. It is clear that when the Knudsen number  is large, $e$ goes to zero so that the error decays quickly. This means that CIS is very efficient for highly rarefied gas flows, i.e., the converged solution can be found within dozens of iterations~\cite{SuArXiv2019}. On the contrary, $e\rightarrow1$ when $\text{Kn}\rightarrow0$, which means that the CIS is extremely slow in the (near) continuum flow regime.


\section{The scheme II GSIS}\label{section_GSIS2}

To expedite the convergence of inner iteration, that is, to reduce the number of $k$ in Eq.~\eqref{Shakhov_time_dependent}, macroscopic synthetic equations are needed. There are several ways of constructing these equations, and we first develop the GSIS-II for time-dependent problems due to its relative simplicity~\cite{Zhu2021JCP}. 

First, in GSIS, given the value of velocity distribution function $h_{n+1}^{k}$ at the $k$-th iteration step, its value at the intermediate $(k+1/2)$-th step is obtained in a similar way to Eq.~\eqref{Shakhov_time_dependent}:
\begin{equation}\label{Shakhov_time_dependent_intermediate}
\frac{h^{k+1/2}_{n+1}-h_n}{\Delta{t}}+
\frac{\bm{v} }{2}\cdot
\left( \frac{\partial{h^{k+1/2}_{n+1}}}{\partial\bm{x}}
+ \frac{\partial{h_{n}}}{\partial\bm{x}}\right) =r\left({\mathcal{L}_{n+1}^{+,k}- \delta_{rp}  h_{n+1}^{k+1/2}}\right)
+(1-r){ \mathcal{L}_{n} }. 
\end{equation}
This velocity distribution function $h^{k+1/2}_{n+1}$ will be used to construct high-order constitutive relations in macroscopic synthetic equations; and when the synthetic equations are solved to obtain macroscopic quantities, say, $M^{k+1}_{n+1}=[\varrho,\bm{u}, \tau,\bm{q}]$,  they will be used in the gain term~\eqref{LBE_Shakhov} for the next inner iteration, until convergence criterion is met. 

Certainly, the macroscopic synthetic equations should be derived exactly from the gas kinetic model. In GSIS-II, the constitutive relations are constructed, with a free parameter $\delta$, in the following manner~\cite{Zhu2021JCP}:
\begin{eqnarray}
\sigma^{k+1}_{ij} =-2\delta^{-1}\frac{\partial u^{k+1}_{<i}}{\partial {x_{j>}}}
+\left(\sigma^{k+1/2}_{ij} +2\delta^{-1}\frac{\partial u^{k+1/2}_{<i}}{\partial {x_{j>}}}\right), \label{sigma_HoT}\\
q^{k+1}_i =-\frac{5}{4\mathrm{Pr}}\delta^{-1} \frac{\partial \tau^{k+1}}{\partial x_i}+
\left(q_i^{k+1/2} +\frac{5}{4\mathrm{Pr}}\delta^{-1} \frac{\partial \tau^{k+1/2}}{\partial x_i}\right). \label{q_HoT}
\end{eqnarray}

At first sight it seems that the GSIS-II is very similar to the HOLO~\cite{Taitano2014}. However, in HOLO the effective rarefaction parameter $\delta$ is chosen to be infinity in Eqs.~\eqref{sigma_HoT} and \eqref{q_HoT}, while in GSIS the NSF constitution relations are explicitly included in the macroscopic synthetic equations by setting $\delta=\delta_{rp}$, which allows fast convergence and unconditional stability when using large time step, as will be shown later.


The macroscopic synthetic equations are solved by the following Crank-Nicolson scheme (although other schemes can also be used):
\begin{equation}\label{eq123_lin_time_dependent}
\begin{aligned}[c]
\frac{\varrho^{k+1}}{\Delta{t}}
+\frac{1}{2}\frac{\partial {u^{k+1}_i}}{\partial{x_i}}=&\frac{\varrho_n}{\Delta{}t_n}
-\frac{1}{2}\frac{\partial {u_{i,n}}}{\partial{x_i}},\\
2\frac{u_i^{k+1}}{\Delta{t}}
+\frac{1}{2}\frac{\partial }{\partial{x_i}}
\left(
\varrho^{k+1}+ {\tau^{k+1}}+\sigma^{k+1}_{ij}\right)
=&
2\frac{u_{i,n}}{\Delta{t}}
-\frac{1}{2}\frac{\partial }{\partial{x_i}}\left(
{\varrho_n}+\tau_n+\sigma_{ij,n}\right), \\
\frac{3}{2}\frac{\tau^{k+1}}{\Delta{t}}
+\frac{1}{2}\frac{\partial {{q^{k+1}_{i}}}}{\partial{x_i}}
+\frac{1}{2}\frac{\partial {{u^{k+1}_{i}}}}{\partial{x_i}}=
&\frac{3}{2}\frac{\tau_n}{\Delta{t}}
-\frac{1}{2}\frac{\partial {q_{i,n}}}{\partial{x_i}}
-\frac{1}{2}\frac{\partial {u_{i,n}}}{\partial{x_i}},	
\end{aligned}
\end{equation}
where terms in the left-hand-side are evaluated at the $(n+1)$-th time step (for clarity the subscript is ignored), while these on the right-hand-side are evaluated at the $n$-th time step.


Therefore,
to calculate the convergence rate of the time-dependent GSIS, the error functions are defined as
\begin{equation}\label{Y_ansatz2_time_dependent} 
\begin{aligned}[c]
Y^{k+1/2}(\bm{x},\bm{{v}}\,)=h_{n+1}^{k+1/2}(\bm{x},\bm{{v}}\,)-h_{n+1}^{k}(\bm{x},\bm{{v}}\,)=e^{k}\bar{Y}(\bm{{v}}\,)\exp(i\bm{\theta}\cdot{\bm{x}}\,),\\
\Phi^{k+1}(\bm{x}\,)=M_{n+1}^{k+1}(\bm{x}\,)-M_{n+1}^{k}(\bm{x}\,)=e^{k+1}\alpha\exp(i\bm{\theta}\cdot{\bm{x}}).
\end{aligned}
\end{equation}
Note that here the macroscopic quantities $M^{k+1}$ are calculated from the synthetic equations~\eqref{eq123_lin_time_dependent}, rather than directly from the velocity distribution function $h^{k+1/2}$. After some algebra, the error decay rate can be obtained by solving the following linear systems:
\begin{equation}\label{L_lin2_time_dependent}
\begin{aligned}[c]
e\left[\frac{2}{\Delta{t}}\alpha_\varrho+i\theta_1\alpha_{u_1}+i\theta_2\alpha_{u_2}+i\theta_3\alpha_{u_3}\right]=0, \\
e\left[\left( \frac{4}{\Delta{t}}+\frac{{\theta}^2}{\delta} \right)\alpha_{u_j}+i\theta_j(\alpha_\varrho+\alpha_{\tau})\right]=S_{j+1}, \\
e\left[\frac{3}{\Delta{t}}\alpha_{\tau}
+i\theta_1(\alpha_{q_1}+\alpha_{u_1})
+i\theta_2(\alpha_{q_2}+\alpha_{u_2})
+i\theta_3(\alpha_{q_3}+\alpha_{u_3})\right]=0,\\
e\left(\frac{5i}{4\text{Pr}}\theta_j{\delta^{-1}}\alpha_{\tau}+\alpha_{q_1}\right)=S_{j+5},
\end{aligned}
\end{equation}
where $j=1,2,3$, and the source terms are 
\begin{equation}\label{L_lin_shcemeII}
\begin{aligned}[c]
S_2=\int\left[{\delta^{-1}}{\theta}^2{}{v}_1
-2i\theta_1v_{\langle{1}}v_{1\rangle}-2i\theta_2{v}_1{v}_2-2i\theta_3{v}_1{v}_3\right]\bar{Y}(\bm{{v}}\,)\mathrm{d}^3\bm{{v}}, \\
S_3=\int\left[{\delta^{-1}}{\theta}^2{}{v}_2
-2i\theta_2v_{\langle{2}}v_{2\rangle}-2i\theta_1{v}_1{v}_2-2i\theta_3{v}_2{v}_3\right]\bar{Y}(\bm{{v}}\,)\mathrm{d}^3\bm{{v}}, \\
S_4=\int\left[{\delta^{-1}}{\theta}^2{}{v}_3
-2i\theta_3v_{\langle{3}}v_{3\rangle}-2i\theta_1{v}_1{v}_3-2i\theta_2{v}_2{v}_3\right]\bar{Y}(\bm{{v}}\,)\mathrm{d}^3\bm{{v}}, \\
S_{j+5}=\int\left[\frac{5i}{4\text{Pr}}\theta_j{\delta^{-1}}{}\left(\frac{2}{3}{v}^2-1\right)+{v}_j\left({v}^2-\frac{5}{2}\right)\right]\bar{Y}(\bm{{v}}\,)\mathrm{d}^3\bm{{v}}.
\end{aligned} 
\end{equation}

Equations~\eqref{L_lin2_time_dependent} and~\eqref{L_lin_shcemeII} can be rearranged as
$L_8e\alpha_M^\top=R_8\alpha_M^\top$,
and the error decay rate can be obtained from the eigenvalues of the matrix $L_8^{-1}R_8$. A comparison of GSIS-II and HOLO is shown in Fig.~\ref{fig:GSIS12_time_dependent}, which clearly show that HOLO is unstable at large time step, say, when $\Delta{t}=15$ and $|\bm{\theta}|=1$. When the perturbation wavevector $\theta$ in Eq.~\eqref{ansatz} is increased, the time step for stable iteration in HOLO is much reduced, while the GSIS is always stable, see the last subfigure. 

\section{The scheme I GSIS}\label{section_GSIS1}

This time-dependent scheme is modified from  Ref.~\cite{SuArXiv2019} which is initially developed to find steady-state solutions of the Boltzmann equation and kinetic model equations. In addition the the five macroscopic equations from the mass, momentum and energy consecration~\eqref{eq123}, evolution equations for the stress and heat flux are also included, like that in the Grad 13 moment equations~\cite{Grad1949}. That is,  we multiply Eq.~\eqref{bgkfd} by $v_{\langle{i}}v_{j\rangle}$ and integrate the resultant equation with respect to the molecular velocity $\bm{v}$, and obtain 
\begin{equation}\label{HoT_sigma0}
\frac{\partial \sigma_{ij}}{\partial {t}}
+2\int{v_{\langle{i}}v_{j\rangle}} \bm{v}\cdot\frac{\partial h}{\partial \bm{x}}d\bm{v}=-\delta_{rp}\sigma_{ij}.
\end{equation}
However, unlike the Grad 13 moment system where the high order term (i.e. the second term in the left-hand-side) is closed by expanding the velocity distribution function into Hermite polynomial of molecular velocity, in GSIS-I Eq.~\eqref{HoT_sigma0} is rearranged in the following form to reflect Newton's law for shear viscosity, which shall allow numerical stability at large time steps:
\begin{equation}\label{HoT_sigma}
\begin{aligned}[c]
\frac{\partial \sigma_{ij}}{\partial {t}}
&+\underbrace{2\int{v_{\langle{i}}v_{j\rangle}} \bm{v}\cdot\frac{\partial h}{\partial \bm{x}}d\bm{v}-2\frac{\partial{u_{<i}}}{\partial {x_{j>}}}}_{\text{HoT}_{\sigma_{ij}}}
+\underbrace{2\frac{\partial{u_{<i}}}{\partial {x_{j>}}}=-\delta_{rp}\sigma_{ij}}_{\text{Newton's law}}.
\end{aligned}
\end{equation}
This equation is solved, again, by the Crank-Nicolson scheme:
\begin{equation}\label{HoT_sigma2}
\begin{aligned}[c]
\left(\frac{1}{\Delta{t}}
+\frac{\delta_{rp}}{2}\right)
&\sigma^{k+1}_{ij}
+\frac{\partial{u^{k+1}_{<i}}}{\partial {x_{j>}}}
\\
&=\left[\left(\frac{1}{\Delta{t}}-\frac{\delta_{rp}}{2}\right)\sigma_{ij,n}
-\frac{\partial{u_{<i,n}}}{\partial {x_{j>}}}-\frac{\text{HoT}_{\sigma_{ij},n}}{2}
\right]
-\frac{\text{HoT}^{k+1/2}_{\sigma_{ij}}}{2}.
\end{aligned}
\end{equation}
It is noted that, when the inner iteration converges,  $\text{HoT}^{k+1/2}_{\sigma_{ij}}$ will be the same as $\text{HoT}^{k+1}_{\sigma_{ij}}$, and hence this numerical scheme is a Crank-Nicolson scheme with second-order accuracy in time.

Likewise, we multiply Eq.~\eqref{bgkfd} by $v_i(v^2-5/2)$ and integrate the resultant equation with respect to $\bm{v}$; we obtain 
\begin{equation}\label{HoT_q}
\begin{aligned}
\frac{\partial q_{i}}{\partial {t}}
&+\underbrace{ \int{\left(v^2-\frac{5}{2}\right)}v_i \bm{v}\cdot\frac{\partial h}{\partial \bm{x}}d\bm{v}
	-\frac{5}{4}\frac{\partial{\tau}}{\partial {x_{i}}
} }_{\text{HoT}_{q_i}} 
+\underbrace{\frac{5}{4}\frac{\partial{\tau}}{\partial {x_{i}}}=-\delta_{rp}\text{Pr}q_{i}}_{\text{ Fourier's law}},
\end{aligned}
\end{equation}
which can also be solved by the Crank-Nicolson scheme.


Although the macroscopic synthetic equations~\eqref{eq123}, \eqref{HoT_sigma} and~\eqref{HoT_q} resemble the Grad 13 moment equations~\cite{Grad1949,henning}, no approximations are introduced here since the higher-order terms are computed directly from the velocity distribution function.




According to the Fourier stability analysis, the error decay rate can be obtained by solving the following equations:
\begin{equation}\label{L_lin1_time_dependent}
\begin{aligned}[c]
e\left[\frac{2}{\Delta{t}}\alpha_\varrho+i\theta_1\alpha_{u_1}+i\theta_2\alpha_{u_2}+i\theta_3\alpha_{u_3}\right]=0, \\
e\left[\left(\frac{4}{\Delta{t}}+\frac{{\theta}^2}{\delta_{rp}+\frac{2}{\Delta{t}}}\right)\alpha_{u_j}+i\theta_j(\alpha_\varrho+\alpha_{\tau})\right]=S_{j+1},\\
e\left[\frac{3}{\Delta{t}}\alpha_{\tau}
+i\theta_1(\alpha_{q_1}+\alpha_{u_1})
+i\theta_2(\alpha_{q_2}+\alpha_{u_2})
+i\theta_3(\alpha_{q_3}+\alpha_{u_3})\right]=0,\\
e\left(\frac{5i\theta_j\alpha_{\tau}}{4(\delta_{rp}\text{Pr}+\frac{2}{\Delta{}t})}+\alpha_{q_j}\right)=S_{j+5}
\end{aligned}
\end{equation}
where $j=1,2,3$, and the source terms are
\begin{equation}\label{L_lin2}
\begin{aligned}[c]
S_2=\frac{1}{\delta_{rp}+\frac{2}{\Delta{t}}}
\int\left[{\theta}^2{v}_1-{2\Theta}
(\theta_1v_{\langle{1}}v_{1\rangle}+\theta_2{v}_1 {v}_2+\theta_3{v}_1 {v}_3)\right]\bar{Y}(\bm{{v}}\,)\mathrm{d}^3\bm{{v}}, \\
S_3=\frac{1}{\delta_{rp}+\frac{2}{\Delta{t}}}
\int\left[{\theta}^2{v}_2-{2\Theta}
 (\theta_1{v}_1 {v}_2+\theta_2v_{\langle{2}}v_{2\rangle}+\theta_3{v}_2 {v}_3)\right]\bar{Y}(\bm{{v}}\,)\mathrm{d}^3\bm{{v}}, \\
S_4=\frac{1}{\delta_{rp}+\frac{2}{\Delta{t}}}
\int\left[{\theta}^2{v}_3-{2\Theta}
 (\theta_1{v}_1 {v}_3+\theta_2{v}_2 {v}_3+\theta_3v_{\langle{3}}v_{3\rangle})\right]\bar{Y}(\bm{{v}}\,)\mathrm{d}^3\bm{{v}}, \\
S_{j+5}=\frac{i}{\delta_{rp}\text{Pr}+\frac{2}{\Delta{t}}}
\int\left[\frac{5}{4}\theta_j\left(\frac{2}{3}{v}^2-1\right)-{\Theta}
{v}_j\left({v}^2-\frac{5}{2}\right)\right]\bar{Y}(\bm{{v}}\,)\mathrm{d}^3\bm{{v}}.
\end{aligned} 
\end{equation} 
and $\Theta=\theta_1{v}_1+\theta_2v_2+\theta_3v_3$.

Numerical results for the BGK and Shakhov models when $\Delta{t}$=1 and 15 are shown in Fig.~\ref{fig:GSIS12_time_dependent}. For the Shakhov kinetic model, GSIS-I has a smaller error decay rate than GSIS-II, especially when $\text{Kn}\rightarrow0$ the error decay rate of GSIS-I goes to zero, while that of GSIS-II goes to $1/3$. This is because the heat flux appears in the gain term~\eqref{LBE_Shakhov} of the Shakhov model, and the GSIS-I has the synthetic equation~\eqref{HoT_q} to guild the evolution of heat flux. However, GSIS-II does not have this capability, which results in a slower convergence  (larger error decay rate $e$) than GSIS-I. For the BGK model, both GSIS schemes have the error decay rate approaching zero when $\text{Kn}\rightarrow0$, because only the density, velocity and temperature appears in the collision term, and both GSIS schemes have the evolution equations for these macroscopic quantities.

\section{The property of asymptotic preserving}\label{section_AP}
    
%

From a practical point of view, the property of  asymptotic NSF preserving should be investigated based on the numerical scale solving the kinetic equation and macroscopic synthetic equations~\cite{Guo2019UP_arXiv}. Since in GSIS the kinetic equation and synthetic equations can be solved by different numerical methods with different order of accuracy, in this section, we consider the influence of spatial and temporal discretizations in the gas kinetic solver on the accuracy of GSIS, based on the assumptions that the spatial grid size $\Delta{x}$ and the time step $\Delta{t}$ are refined enough to capture the physical solution of NSF equations. Namely, we investigate at what values of $\alpha$ and $\beta$, can the macroscopic synthetic equations be exactly reduced to the NSF equations when $\text{Kn}$ is small, through the Chapman-Enskog expansion~\cite{CE} of the discretized gas kinetic equation, with the following scaling:
\begin{equation}\label{scaling}
\begin{aligned}[c]
{\Delta{x}}\sim{\text{Kn}^{1/\alpha}},\quad
{\Delta{t}}\sim{\text{Kn}^{1/\beta}}.
\end{aligned}
\end{equation}
Note that the spatial grid size $\Delta{x}$ and time step $\Delta{t}$ have been normalized by the characteristic flow length $L$ and time ($L/v_m$), respectively. 
Here $\alpha$ and $\beta$ denote the order of accuracy in the asymptotic preserving of NSF equations. Clearly, the larger the values of $\alpha$ and $\beta$, the better the numerical scheme. If $\alpha=\infty$, the scheme will capture the hydrodynamical behavior when $\Delta{x}$ is approximately the system size (no matter what the value of $\text{Kn}$ is), as long as this size is adequate to capture the flow physics.  On the other hand, if  $\beta=\infty$, the scheme will capture the hydrodynamical behavior when $\Delta{t}$ is approximately the characteristic time of the system (e.g., the oscillation period of a sound wave).

Since when the Knudsen number is small, the error decay rate of GSIS is much smaller than unity (i.e., zero in GSIS-I for both BGK and Shakhov models, and 1/3 in GSIS-II scheme when the Shakhov model is used), the converged solution can be found within a few  inner iterations. Thus, we have $h_{eq}^{k+1}=h_{eq}^{k}$ and $M^{k+1}=M^{k}$. When the inner iteration is converged, the iterative scheme~\eqref{Shakhov_time_dependent} can be expressed as
\begin{equation}\label{LBE_GSIS}
\frac{\partial h}{\partial t}
+\bm{v}\cdot\frac{\partial{h}}{\partial\bm{x}}
+O(\Delta{t}^m)\delta_t(h)
+O(\Delta{x}^n)\delta_x(h)
=\mathcal{L}_s,
\end{equation}
where $n$ is the order of approximation for the spatial derivative in the kinetic equation,  while $\delta_x(h)$ is the $(n+1)$-th order derivative of $h$ with respect to spatial coordinates. For instance, if the second-order upwind finite difference scheme, we have $n=2$. 
Similar applies to $m$ and $\delta_t(h)$. 

\subsection{Chapman-Enskog expansion}

In the Chapman-Enskog expansion the velocity distribution function is approximated by the Taylor expansion $h=h^{(0)}+\text{Kn}h^{(1)}+\text{Kn}^2h^{(2)}+\cdots$,
 so are the stress $\sigma_{ij} =\sum_{\ell=0}^\infty \text{Kn}^{\ell} \sigma_{ij}^{(\ell)}$ and heat flux $\bm{q} =\sum_{\ell=0}^\infty \text{Kn}^{\ell} \bm{q}^{(\ell)}$,
where $\sigma^{(\ell)}_{ij}=2\int{v_{\langle{i}}v_{j\rangle}h^{(\ell)}}d\bm{v}$ and
$\bm{q}^{(\ell)}=\int\left({v^2}-\frac{5}{2}\right)\bm{v}h^{(\ell)}d\bm{v}$. However, the five conservative variables $C_M=\{\rho,\bm{u}, \tau\}$  are calculated only according to the zeroth-order expansion. That is, 
\begin{eqnarray}\label{density_hilbert}
\rho=\int{}h^{(0)}d\bm{v}, \quad \bm{u}=\int{}\bm{v}h^{(0)}d\bm{v}, \quad \tau=\frac{2}{3}\int{}\left(v^2-\frac{3}{2}\right)h^{(0)}d\bm{v},
\end{eqnarray}
with the compatibility condition  $\int{}h^{(\ell)}d\bm{v}=\int{}\bm{v}h^{(\ell)}d\bm{v}=\int{}v^2h^{(\ell)}d\bm{v}=0$ for $\ell\ge1$.
From Eq.~\eqref{density_hilbert} and the compatibility condition, one finds that the time derivatives in Eq.~\eqref{eq123}  can be formally written as a series in $\text{Kn}$~\cite{henning}:
$\frac{\partial}{\partial{} t}=\sum_{\ell=0}^\infty \text{Kn}^{\ell} \frac{\partial}{\partial{} t_{\ell}}$.

%

\subsection{GSIS-I}

By substituting  the Taylor expansion of velocity distribution function into Eq.~\eqref{LBE_GSIS} and collecting terms with the order of $\text{Kn}^{-1}$, we have 
\begin{equation}\label{feq_zero}
h^{(0)}=\left[\varrho+2\bm{u}\cdot\bm{v}+\tau\left(v^2-\frac{3}{2}\right)\right]f_{eq},
\end{equation}
and $\sigma^{(0)}_{ij}=\bm{q}^{(0)}=0$, with the following largest scaling:
\begin{equation}\label{scaling1}
\begin{aligned}[c]
\Delta{x}\sim{\text{Kn}^{1/\infty}}=O(1),\quad
\Delta{t}\sim{}{\text{Kn}^{1/\infty}}=O(1).
\end{aligned}
\end{equation}

Under this circumstance, by collecting terms with the order $\text{Kn}^{0}$, we have
\begin{equation}\label{h1_taylor}
h^{(1)}=-\frac{\partial h^{(0)}}{\partial t_0}-\bm{v}\cdot\frac{\partial{h^{(0)}}}{\partial\bm{x}}
\underbrace{-O(\Delta{t}^m)\delta_t(h)
-O(\Delta{x}^n)\delta_x(h)}_{\text{discretization error terms}}.
\end{equation}
Note that in the standard Chapman-Enskog expansion, this $h^{(1)}$ term is used to produce the NSF constitutive relations~\eqref{NSF_relations_2}. Although there are some error terms in $h^{(1)}$, this does not affect the exact derivation of NSF constitutive relations in GSIS-I. This is because we only need $h^{(0)}$ to evaluate the stress and heat flux according to Eqs.~\eqref{HoT_sigma} and~\eqref{HoT_q}, while the $h^{(1)}$ term will result constitutive relations at the order of $\text{Kn}^2$. To be specific, let us take the stress for an example. When the inner iteration is converged, Eq.~\eqref{HoT_sigma}
becomes
\begin{equation}\label{HoT_sigma_section5}
\begin{aligned}[c]
\frac{\partial \sigma_{ij}}{\partial {t}}
&+2\int{v_{\langle{i}}v_{j\rangle}} \bm{v}\cdot\frac{\partial h}{\partial \bm{x}}d\bm{v}=-\delta_{rp}\sigma_{ij}.
\end{aligned}
\end{equation}
Since $\sigma_{ij}\propto\text{Kn}$, the leading order solution is
\begin{equation}
\begin{aligned}[c]
\sigma^{(1)}_{ij}&=-\frac{2}{\delta_{rp}}\int{v_{\langle{i}}v_{j\rangle}} \bm{v}\cdot\frac{\partial h^{(0)}}{\partial \bm{x}}d\bm{v}
=-\frac{2}{\delta_{rp}}\frac{\partial u_{<i}}{\partial {x_{j>}}},
\end{aligned}
\end{equation}
which is exactly Newton's law of stress. 

%

Therefore, GSIS-I asymptotically preserves the NSF equations with the largest scalings~\eqref{scaling1}. That is to say, as long as the spatial resolution $\Delta{x}=O(1)$ and temporal resolution $\Delta{t}=O(1)$ are able to capture the physical solution of NSF equations, GSIS-I is able to recover the linearized NSF equations when $\text{Kn}\rightarrow0$. This means that the overall order of accuracy of GSIS-I depends only on the order of accuracy in solving macroscopic synthetic equations.
In reality, however, such a large spatial grid size $\Delta{x}=O(1)$ cannot be used in regions with Knudsen layer or shock structure, where the physical solutions require a spatial resolution of $O(\text{Kn})$. Fortunately, these kinetic layers only take up a small fraction of the computational domain, say, in the vicinity of solid walls, which can be captured by implicit schemes with non-uniform spatial discretization. This will be demonstrated in section~\ref{Osci_Couette} below. Likewise, such a large temporal step $\Delta{t}=O(1)$ will be much reduced to the maximum time step in solving the NSF equations accurately; this will be demonstrated in section~\ref{Osci_CRBS}.

\subsection{GSIS-II}

We now follow the standard procedure to see at what conditions can the NSF equations be recovered from the macroscopic synthetic equations in GSIS-II. At the zeroth-order approximation, we have Eq.~\eqref{feq_zero}, and the following Euler equation:
	\begin{equation}\label{eq123_0}
	\begin{aligned}
	\frac{\partial {\varrho}}{\partial{t_0}}+\frac{\partial {u_i}}{\partial{x_i}}=0, \quad
	2\frac{\partial {u_i}}{\partial{t_0}}+\frac{\partial {\varrho}}{\partial{x_i}}+\frac{\partial {\tau}}{\partial{x_i}}=0, \quad
	\frac{3}{2}\frac{\partial {\tau}}{\partial{t_0}}+\frac{\partial {u_j}}{\partial{x_j}}=0,
	\end{aligned}
	\end{equation}
	that relates the time derivative to spatial derivatives.

With the largest scaling~\eqref{scaling1}  $h^{(1)}$ is given by Eq.~\eqref{h1_taylor}. If one calculates the stress and heat flux according to Eq.~\eqref{MP}, then the constitutive relations are not exactly the NSF ones due to the presence of error terms. That is to say, in general, the NSF equations cannot be exactly recovered at such large temporal step and spatial size. In practice, if the error terns are very small, GSIS-II can also approach the NSF limit in the continuum flow regime. This will be demonstrated in section~\ref{Taylor_vortex}.

In GSIS-II, in order to recover the NSF constitutive relations~\eqref{NSF_relations_2} exactly in general cases, 
the velocity distribution function in the Taylor expansion to the first-order of Knudsen number must be exactly recovered as 
\begin{equation}\label{h1_exact}
\begin{aligned}[c]
h^{(1)}=&-\frac{\partial h^{(0)}}{\partial t_0}
-\bm{v}\cdot\frac{\partial{h^{(0)}}}{\partial\bm{x}}=-\left[
2v_{\langle{i}}v_{j\rangle}
\frac{\partial u_{\langle{i}}}{\partial x_{j\rangle}}
+\left(v^2-\frac{5}{2}\right)v_i\frac{\partial \ln\tau}{\partial x_i}
\right]f_{eq}.
\end{aligned}
\end{equation}
where $ h^{(0)}$ is given in Eq.~\eqref{feq_zero}, and the time derivative is changed to spatial derivatives with the help of Eq.~\eqref{eq123_0}. This requires the following scaling
\begin{equation}\label{scaling2}
\begin{aligned}[c]
\Delta{x}\sim{\text{Kn}^{1/n}},\quad
\Delta{t}\sim{\text{Kn}^{1/m}}.
\end{aligned}
\end{equation}

\begin{figure}[t]
	\centering
	\includegraphics[scale=0.5]{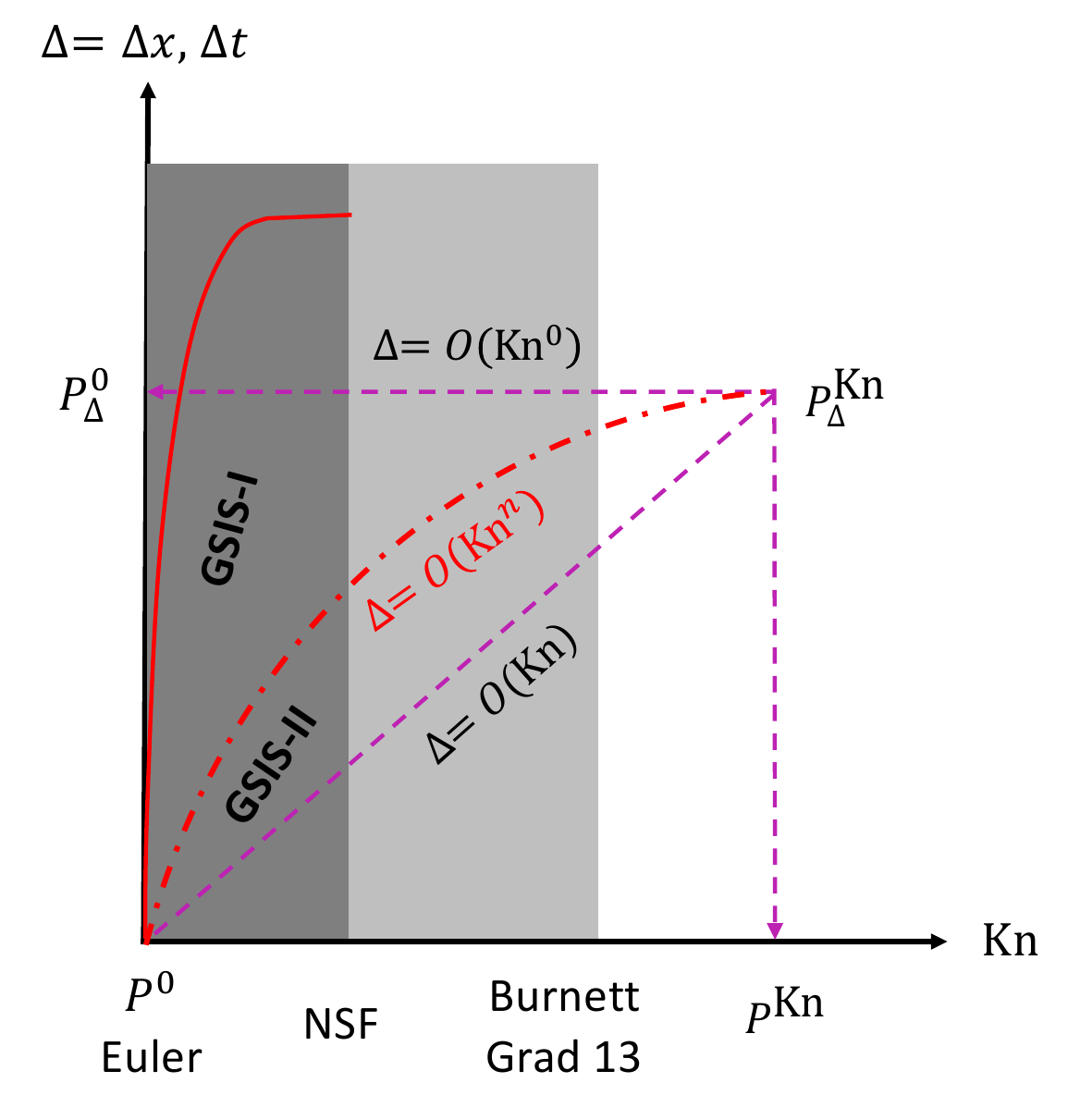}
	\caption{
		Schematic of the asymptotic path to the limiting hydrodynamic flow regimes. The NSF equations are valid in the dark gray region. The region below the line $\Delta=O(\text{Kn})$ suggests resolved kinetic scale. The
		red solid line stands for the maximum spatial grid size/time step to solve the NSF equations accurately using some discretization method (certainly $\Delta$ depends on the numerical method for NSF equations and specific flow problems; $\Delta$ could be larger than $O(1)$ because the time step can be proportional to $\text{Kn}^{-1}$ in some flows), below which GSIS-I can capture the hydrodynamic behavior. GSIS-II works at a smaller value of $\Delta$, say, beneath the dash-dotted line $\Delta=O(\text{Kn}^n)$. Due to the instability at large time step, the validation range of HOLO is smaller than GSIS-II.
	}
	\label{fig:GSIS_UP}
\end{figure}

Normally the kinetic equation is solved with second-order accuracy both in temporal and spatial directions,  that is, $n=m=2$. Therefore, the NSF equations are recovered in GSIS-II with $\Delta{x}\sim\sqrt{\text{Kn}}$ and $\Delta{t}\sim\sqrt{\text{Kn}}$, like the (discrete) unified gas-kinetic scheme~\cite{UGKS2010JCP,guo2013discrete,Guo2019UP_arXiv}.

The asymptotic path to the limiting hydrodynamic flow regimes are summarized in Figure~\ref{fig:GSIS_UP}. 


\section{Numerical results}\label{section_numerical}

In this section we propose several solid numerical examples to assess the accuracy and efficiency of both GSIS schemes, as well as HOLO. 

\subsection{Rayleigh-Brillouin scattering}\label{Osci_CRBS}

 The coherent Rayleigh-Brillouin scattering is a promising technique to probe the property of gas, in which the wavelike density perturbation in gas is created by a moving optical lattice. We choose this problem because it is a zero-dimensional problem so any one can quickly test/compare our and their methods. 
 
 Applying the Fourier transform in the scattering (say, $x_1$) direction, the governing equation for the velocity distribution function $h$ can be written as~\cite{Wu2020AIA}
\begin{equation}\label{CRBS_LBE}
\frac{\partial {h}}{\partial {t}}+2i\pi{}v_1h
=\mathcal{L}(h)
+2v_1\cos(2\pi{f_s}t)f_{eq},
\end{equation}
where $f_s$ is the frequency of the moving optical lattice; we choose $f_s=\sqrt{5/6}$ (i.e. the sound speed normalized by the most probable speed of gas molecules) so that the amplitude of perturbed density will scale as $1/\text{Kn}$ when the Knudsen number is small. If the numerical scheme cannot preserve the NSF limit when $\text{Kn}\rightarrow0$, then a very small time step is needed to keep a low numerical dissipation; otherwise the amplitude will be much smaller than the converged solution.

Since the spatial derivative $\partial/\partial{x_1}$ is replaced by $2i\pi$ in the Fourier transform, the conservation equations~\eqref{eq123} becomes
\begin{eqnarray}
\frac{\partial{\varrho}}{\partial t}+2i\pi{{u}_1}=0, \quad
\frac{3}{2}\frac{\partial{\tau}}{\partial t}+2i\pi(u_1+q_1)=0, \\
2\frac{\partial{u}_1}{\partial t}+2i\pi(\varrho+\tau+\sigma_{11})=2\cos(2\pi{f_s}t),
\end{eqnarray}
and the same transform is applied to the synthetic equations for stress and heat flux in Eqs.~\eqref{sigma_HoT}, \eqref{q_HoT}, \eqref{HoT_sigma}, and \eqref{HoT_q}. 


\begin{figure}[t]
	\centering
	\includegraphics[scale=0.3]{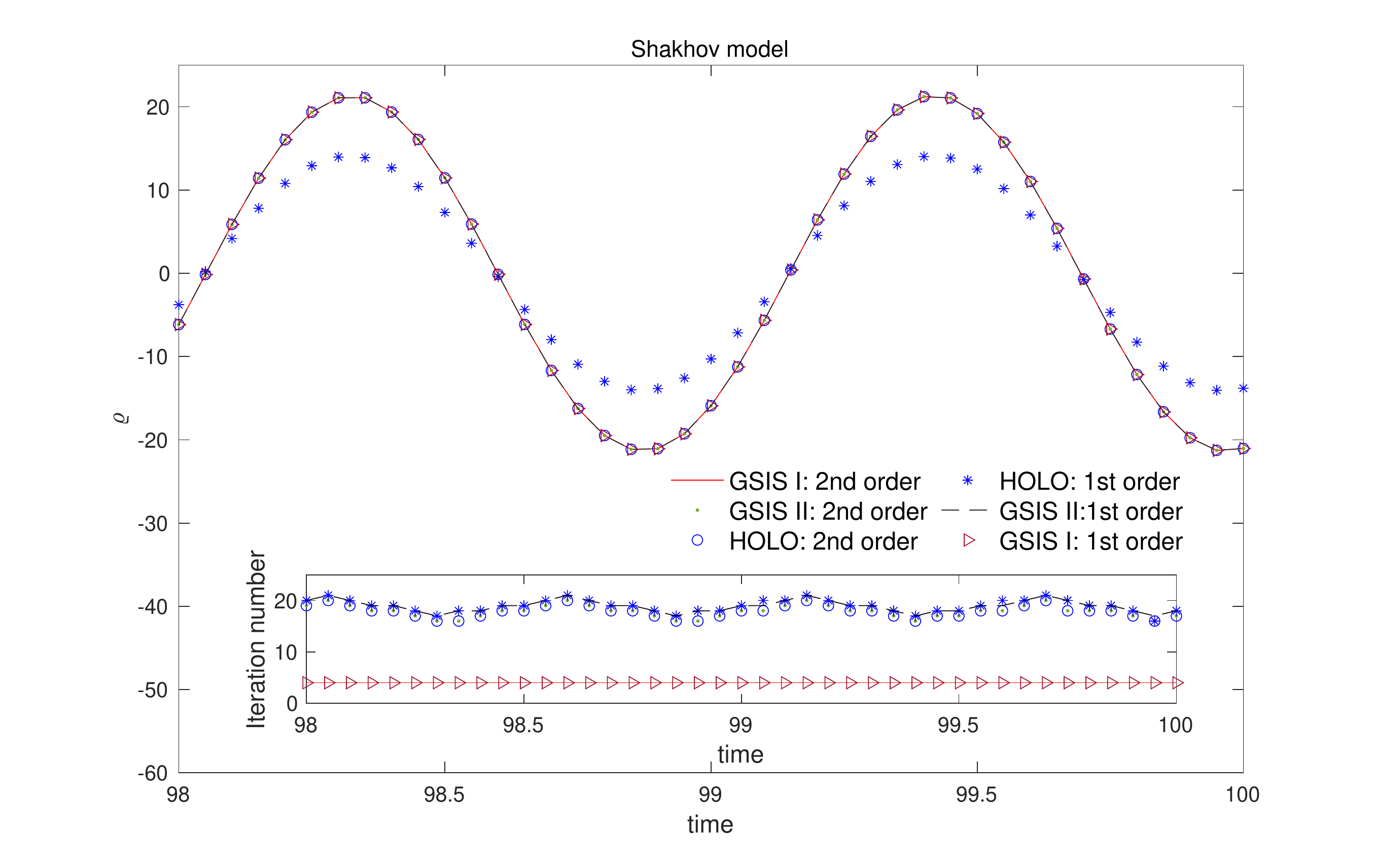}
	\includegraphics[scale=0.3]{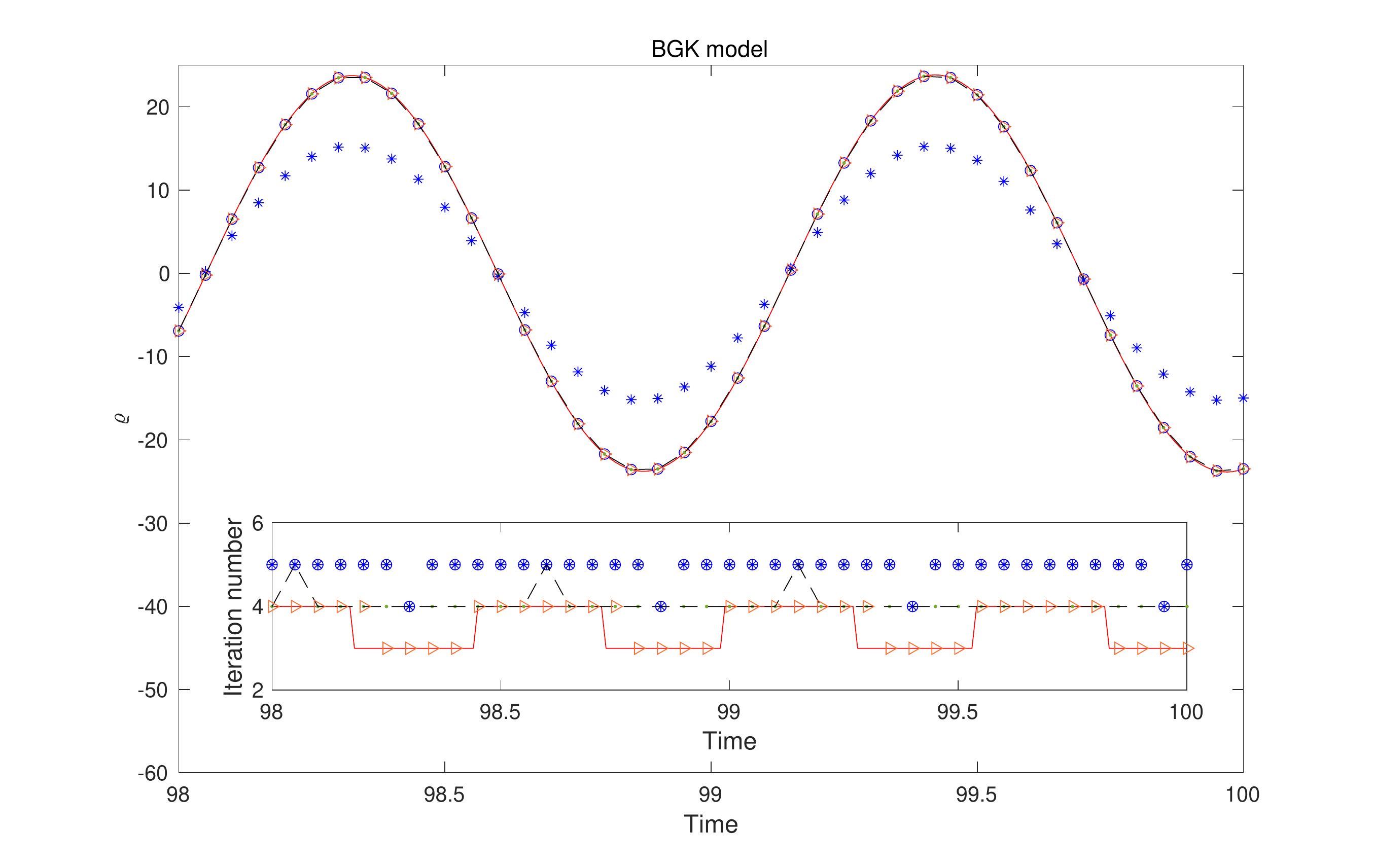}
	\caption{
		The imaginary part of the density perturbation $\varrho$ in the coherent Rayleigh-Brillouin scattering, when the rarefaction parameter is $\delta_{rp}=1000$. Inset shows the iteration number, where the iteration is terminated when the relative error in macroscopic quantities between two consecutive iterations is less than $10^{-10}$. The kinetic equations are solved with a time step of $\Delta{t}=0.01$, by the backward Euler method and Crank-Nicolson scheme, respectively. The macroscopic synthetic equations are always solved by the Crank-Nicolson scheme with the same time step.
	}
	\label{fig:CRBS}
\end{figure}

In the numerical simulation, the molecular velocity space $\bm{v}$ is discretized by $6\times6\times6$ Gauss-Hermite quadrature, which is accurate to resolve the velocity distribution function when the Knudsen number is small: in the continuum flow regime the velocity distribution function contains $v^3_i$, therefore, the integrand in the high order term in Eq.~\eqref{HoT_q} has the highest polynomial of $v^7_i$, while the Gauss-Hermite quadrature of order 6 is accurate for polynomial up to the order of 11. Starting from the zero initial values for the velocity distribution function and macroscopic quantities, the synthetic equations are solved by the Crank-Nicolson scheme, while the kinetic equation is solved by backward Euler and Crank-Nicolson schemes, respectively. The inner iteration terminates when the relative error in density between two consecutive steps are less than $10^{-10}$.

It should be noted that in HOLO, in order to make the mesoscopic and macroscopic equations consistent, consistency terms are introduced to enslave the solution of synthetic equations to that of the kinetic equation~\cite{Taitano2014}. In this problem, since the spatial derivative is handled exactly by the Fourier transform, and the molecular velocity space is discretized by adequate quadrature, these consistency terms vanish. In GSIS, however, we enslave the solution of kinetic equation to that of the synthetic equations, since the kinetic equation converges so slowly that false convergence might happen~\cite{DSA2002,SuArXiv2019}. That is, when the inner iteration is converged (judged by the relative error in macroscopic quantities), the velocity distribution function is updated to reflect the converged macroscopic quantities:
\begin{equation}\label{corretion}
\begin{aligned}
h^{k+1}=h^{k+1/2}
&+ \left({\varrho}^{k+1}-\varrho^{k+1/2}\right)  +2\left({\bm{u}}^{k+1}-\bm{u}^{k+1/2}\right)\cdot\bm{v} \\
&+\left({\tau}^{k+1}-\tau^{k+1/2} \right) \left( v^2-\frac{3}{2} \right).
\end{aligned}
\end{equation}

Numerical results of the GSIS and HOLO for coherent Rayleigh-Brillouin scattering is shown in Fig.~\ref{fig:CRBS}. From the figure we see that, when the kinetic equation is solved with second-order accuracy, GSIS-I, GSIS-II, and HOLO yield identical solutions. For the BGK model, in each time step, CIS needs around 200 inner iterations to find converged solutions (not shown), while GSIS-I, GSIS-II, and HOLO needs about 5 iterations. This is consistent with the Fourier stability analysis presented in Fig.~\ref{fig:GSIS12_time_dependent}, since the wavevector in this problem is always unity. For the Shakhov model, GSIS-I still needs 5 iterations, while GSIS-II and HOLO need about 20 iterations at each time step. This is because the former has synthetic equation for the evolution of heat flux that appears in the gain term of kinetic model so that the error decay rate approaches zero when $\text{Kn}\rightarrow0$, while the latter has the error decay rate approaching $1/3$.

We also find that, when the kinetic equation is solved with first-order temporal accuracy, both GSIS schemes yield accurate solutions. However, HOLO exhibits visible dissipation as the density amplitude is smaller. This is because the time step should be no larger than $O(\text{Kn})$ for first-order scheme, otherwise the artificial viscosity is comparable or larger than the physical viscosity, which leads to inaccurate solutions. It is very interesting to note that, if we turn off the correction of velocity distribution function in Eq.~\eqref{corretion}, both GSIS schemes produce the same  inaccurate result as HOLO. This clearly demonstrates that we should enslave the solution of kinetic equation to that of the synthetic equation, while the HOLO does the reverse way and get inaccurate results.

Finally, it should be noted that although GSIS-I allows a time step of $O(1)$ to asymptotic preserve the NSF limit, here it is chosen as $\Delta{t}=0.01$ because the NSF equations cannot be accurately solved by a second-order scheme when the time step is larger than 0.01 in this specific problem.

\begin{figure}[t]
	\centering
	\includegraphics[scale=0.4]{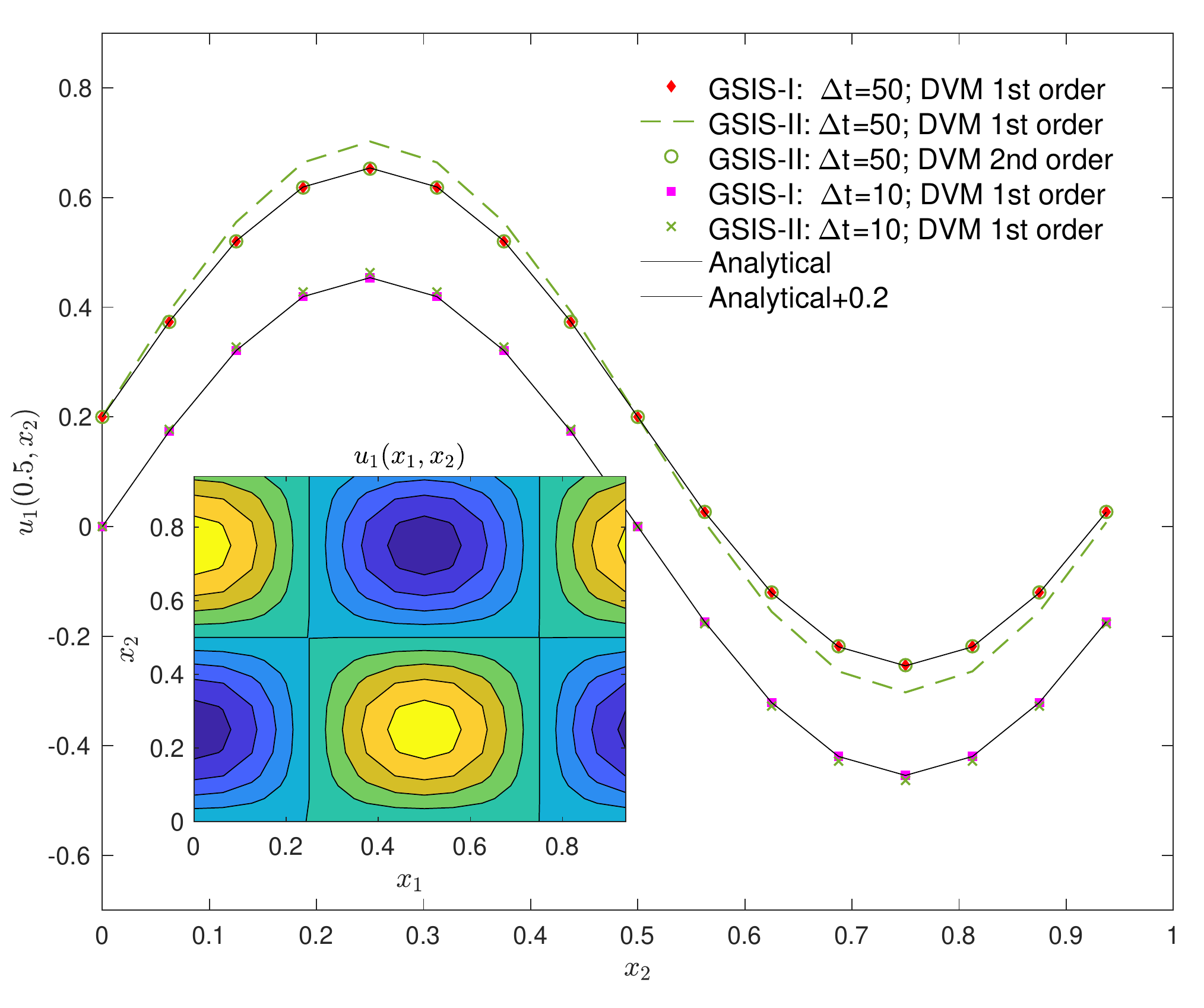}
	\caption{
		Velocity profiles in the two-dimensional Taylor vortex flow at the time $t=200$, when the BGK model with $\delta_{rp}=10000$ is solved by GSIS. The profiles with time step $\Delta{t}=50$ are shifted upward for clarity. DVM 1st (2nd) order means that the kinetic equation is solved with first (second) order temporal accuracy.  Inset: velocity contour at $t=200$.
	}
	\label{fig:Taylor}
\end{figure}

\subsection{Decay of two-dimensional Taylor vortex}\label{Taylor_vortex}

To test the property of fast converging and asymptotic NSF preserving of GSIS and HOLO at the time step $\Delta{t}\sim{O(1)}$, we consider the decay of  two-dimensional incompressible Taylor vortex within a periodic domain $0\le{x_1},x_2\le1$. In the continuum flow regime, the flow is governed by the incompressible Navier-Stokes equations and has the following analytical solution:
\begin{equation}
\begin{aligned}[b]
u_1(x_1,x_2,t)=&-\cos(2\pi{x_1})\sin(2\pi{x_2})\exp\left(-{4\pi^2}t/{\delta_{rp}}\right),\\
u_2(x_1,x_2,t)=&~~~\sin(2\pi{x_1})\cos(2\pi{x_2})\exp\left(-{4\pi^2}t/{\delta_{rp}}\right),
\end{aligned}
\end{equation}
where the density and temperature are always zero. Therefore, the BGK model is used.

This is an ideal test case to assess the asymptotic NSF preserving property, since when $\delta_{rp}$ is large ($\text{Kn}$ is small), the Taylor vortex takes a long time to decay, so the time step can be made very large if the numerical scheme for the kinetic system asymptotic preserves the NSF equations.

Without loss of generality we choose $\delta_{rp}=10^4$, and in order to focus only on the error in temporal discretization, we solve the kinetic equation and macroscopic synthetic equation by the Fourier spectral method in the spatial directions, with the spatial grid size of $\Delta{x}=1/16$. The kinetic equation is solved by the backward Euler and Crank-Nicolson scheme for the first- and second-order temporal accuracy, respectively, with the initial condition
$
h(\bm{v},t=0)=2[v_1u_1(x_1,x_2,0)+v_2u_2(x_1,x_2,0)]f_{eq}$,
while the synthetic equations are solved by the second-order Crank-Nicolson scheme. The molecular velocity space $\bm{v}$ is discretized by $6\times6\times6$ Gauss-Hermite quadrature, which is sufficient accurate to evaluate the high order term in Eq.~\eqref{HoT_sigma}.

Numerical results are summarized in Fig.~\ref{fig:Taylor}. When $\Delta{t}=1$, both GSIS and HOLO yield accurate results (not shown for clarity). However, when $\Delta{t}$ is increased to 10, HOLO becomes unstable, which is consistent with the result from the Fourier stability analysis in Fig.~\ref{fig:GSIS12_time_dependent} (note that in this case the wavevector is $\theta=8\pi^2$).  Both GSIS schemes produce stable results, and converged solution in each time step is found within 4 iterations even when the time step is as large as $\Delta{t}=50$. From the figure one can also find that when the kinetic equation is solved with second-order temporal accuracy, accurate results are obtained. However, when the kinetic equation is solved with first-order temporal accuracy, GSIS-II generates inaccurate solutions when the time step is large, say, $\Delta{t}=50$. This is because GSIS-II do not have the property of asymptotic NSF preserving at large temporal step.

To be specific, we find that in GSIS-II, when the streaming operator is treated exactly by the  Fourier spectral method, the first-order term in the Taylor expansion of velocity distribution function is given by Eq.~\eqref{h1_taylor}.
Suppose $\Delta{t}\sim{O(1)}$, it is estimated that the last error term has a contribution to the shear viscosity at the order of 
\begin{equation}
\Delta{t}\left(\frac{4\pi^2}{\delta_{rp}}\right)^{m+1}
=\underbrace{\left(\Delta{t}\frac{16\pi^4}{\delta_{rp}}\right)}_{O(1)}
\left(\frac{4\pi^2}{\delta_{rp}}\right)^{m-1}
\frac{1}{\delta_{rp}},
\end{equation}
where the underbraced term can be made $O(1)$ since this is largest time step to get the Navier-Stokes equation correct when solved by the second-order temporal scheme.
Therefore, when the kinetic equation is solved by the backward Euler scheme, we have $m=1$ and the error term is at  the same order with the viscosity coefficient, hence the GSIS-II is not accurate at large time step. However, when the kinetic equation is solved by the second-order temporal accuracy, we have $m=2$. Therefore, the error term is much smaller than the viscosity coefficient when the Knudsen number is small. In this case, GSIS-II can yield accurate NSF solutions even at large time steps, as observed in Fig.~\ref{fig:Taylor}.


\subsection{Oscillatory Couette flow}\label{Osci_Couette}

Consider the oscillatory Couette flow between two parallel plates and investigate the combined effect of spatial and temporal discretizations. The plate located at $x_1=0$ oscillates in the $x_2$ direction with the velocity
$u_{wall}=\sin(2\pi{f_s}t)$,
while the other plate at $x_1=1$ is stationary. The rarefaction parameter is chosen to be $\delta_{rp}=1000$, while the oscillation frequency is $2\pi{f_s}=0.1$. This problem can be greatly simplified. In fact, we have $\rho=\tau=0$, and the only the evolution equations for the velocity $u_2$ and stress $\sigma_{12}$ are needed.


\begin{figure}[t]
	\centering
	\includegraphics[scale=0.4]{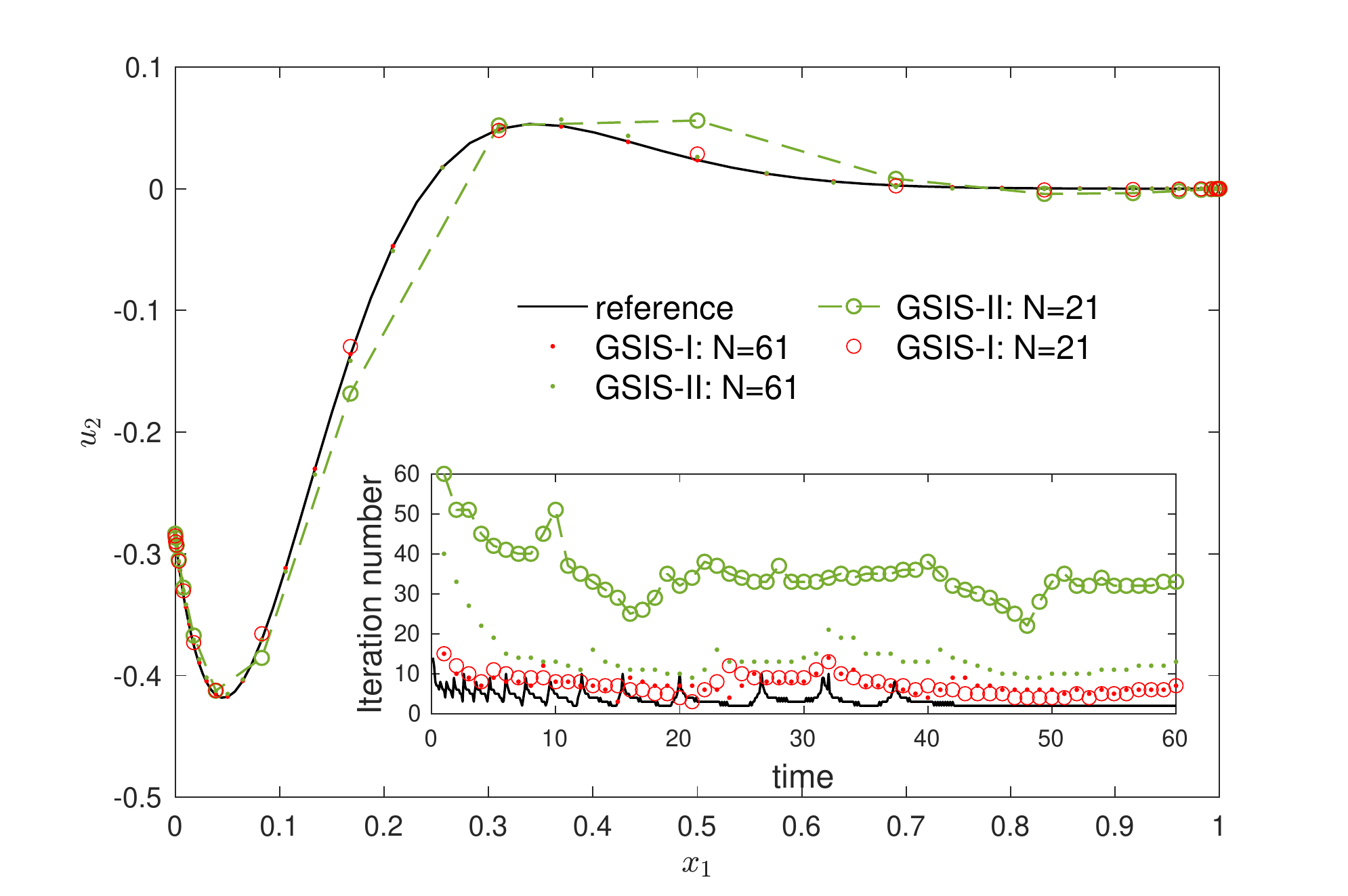}
	\caption{
		Velocity profiles in the oscillating Couette flow at the time $t=60$, when the BGK model with $\delta_{rp}=1000$ is solved by the two GSIS schemes with different spatial resolution, with the time step $\Delta{t}=1$. The reference solution is obtained from GSIS-I with $N=121$ and $\Delta{t}=0.1$. Both kinetic and synthetic equations are solved by the Crank-Nicolson scheme. Inset: the inner iteration number at each time step.  
	}
	\label{fig:Couette}
\end{figure}

The spatial coordinate $x_1\in[0,1]$ is discretized non-uniformly with $N$ points, in the following manner
\begin{equation}\label{spatial_discretization}
x_1=\frac{1}{2}+\frac{\tanh(8m)}{2\tanh(4)}, \quad
j=\frac{0,1,\cdots,N-1}{N-1}-\frac{1}{2},
\end{equation}
so that the Knudsen layer can be well resolved. Later it will be shown this affects the efficiency of inner iterations.  The molecular velocity space $v_2\times{v_3}$ is discretized by $6\times6$ Gauss-Hermite quadrature, while $v_1$ is discretized non-uniformly:
\begin{equation}\label{nonuniform_v}
v_1=\frac{6}{(N_v-1)^3}(-N_v+1,-N_v+3,\cdots,{N_v-1})^3,
\end{equation}
where $N_v=32$ is the total number of discretized velocity in the $v_1$ direction.

The kinetic equation is solved by the Crank-Nicolson scheme with second-order spatial and temporal accuracy, with the initial condition $h(x_1,\bm{v},t=0)=0$, and the boundary conditions are $h(x_1=0,v)=2u_{wall}v_2f_{eq}$ for $v_1>0$  and $h(x_1=1,v)=0$ for $v_1<0$.
The synthetic equations are solved by the second-order temporal accuracy, while the spatial derivative is approximated by central finite-difference with 5 stencils; hence in the solving the synthetic equations $j$ in Eq.~\eqref{spatial_discretization} is taken from 2 to $N-3$, while the 4 left points, obtained from the kinetic equation, provide boundary conditions to synthetic equations.



\begin{figure}[t]
	\centering
	\includegraphics[scale=0.4]{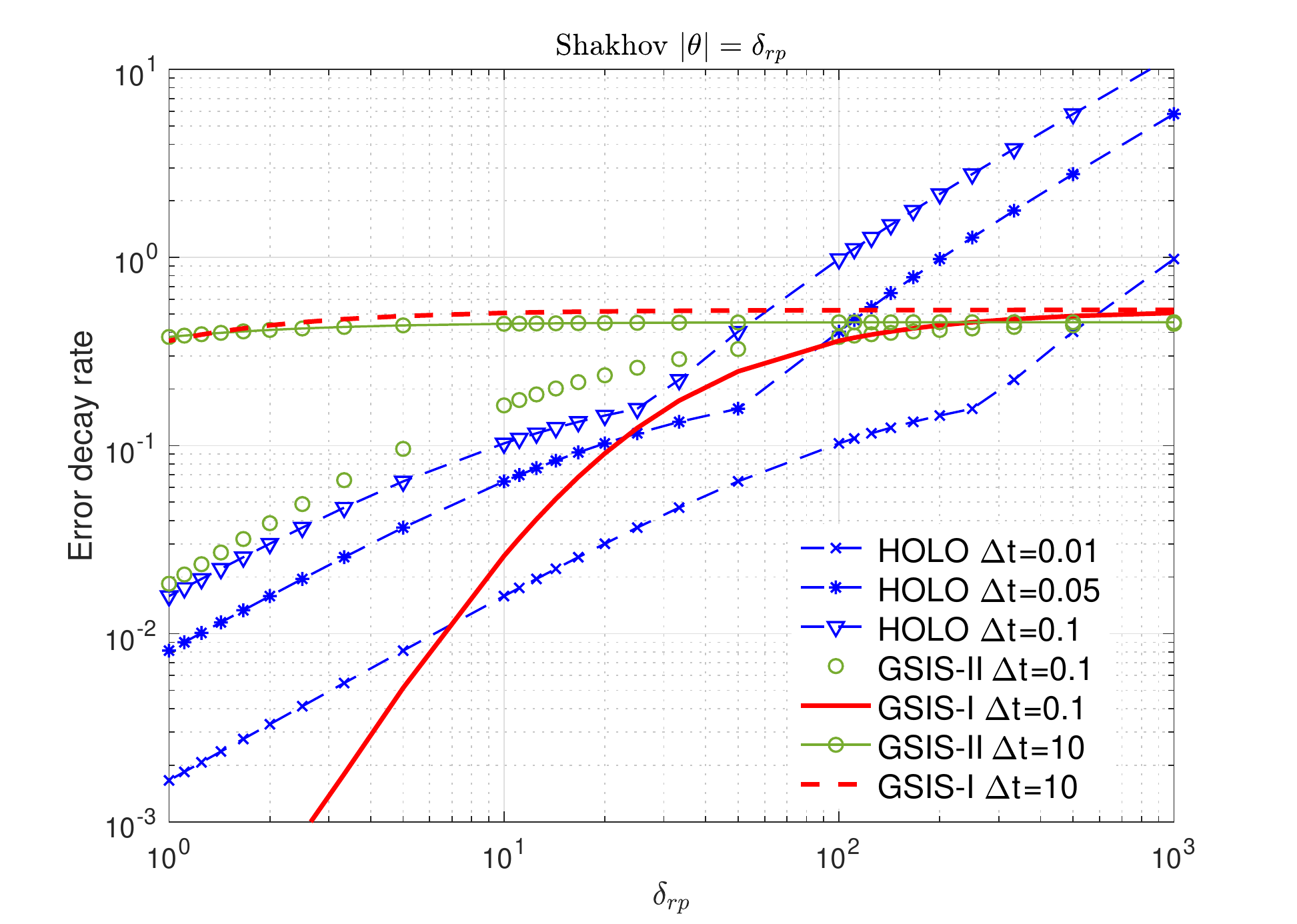}
	\caption{
		The error decay rate as a function of the rarefaction parameter in GSIS and HOLO. For wall bounded problems such as the Poiseuille and Couette flows, the perturbation from the wall bears a wavevector of $|\bm{\theta}|=\delta_{rp}$ if the Knudsen layer is resolved. Therefore, HOLO is only stable when the time step is small, while the robustness of two GSISs is manifested at any time step.
	}
	\label{fig:HOLO_time_dependent}
\end{figure}

Numerical solutions of the oscillatory flow at $\delta_{rp}=1000$ are shown in Fig.~\ref{fig:Couette} for different spatial resolutions. It is seen that when the number of spatial grid is decreased from  $N=121$ to 61 and then to 21, the accuracy of GSIS-I barely reduces. However, in GSIS-II when $N=21$, large difference to the reference solution is observed. This is for the GSIS-II the NSF equations can only be derived exactly when the spatial resolution is about $\Delta{x}={O(\sqrt{\text{Kn}})}\approx0.032$. When $N=121, 61$, and 21, the maximum grid size according to Eq.~\eqref{spatial_discretization} are 0.033, 0.066, and 0.190, respectively. The case of $\max(\Delta{x})=0.190$ is certainly too large.

On the other hand, we find that, when the same spatial resolution is used, generally speaking, GSIS-I needs less iteration numbers than GSIS-II. When $N$ is decreased, the iteration number increases. This is because this flow is driven by the oscillation of left plate; smaller value of $N$ means coarser spatial grid size and hence the information from the plate cannot be effectively passed to the bulk flow regime. Therefore, the use of non-uniform spatial grid~\eqref{spatial_discretization} not only allows the capture of flow dynamics around the Knudsen layer, but also facilitates fast convergence.

Note that the result of HOLO is not shown, since we find that it is not stable when $\Delta{t}>0.004$. This can be understood as follows. When the Knudsen layer is resolved, the oscillating plate will generate a perturbation with a wavelength at the order of mean free path, that is, the wavevector is about $\theta\approx2\pi\delta$. With this information, we re-calculate the error decay rate of GSIS-I, GSIS-II and HOLO in Fig.~\ref{fig:HOLO_time_dependent} with $\theta=\delta_{rp}$. It is seen that GSIS is stable, while HOLO is unstable when $\Delta{t}>0.01$ at $\delta_{rp}=1000$. This is consistent with our numerical observations. If a coarse spatial grid is used, the stability region of HOLO increases, however, the convergence speed is much reduced.

\begin{figure}[t]
	\centering
	\includegraphics[scale=0.37]{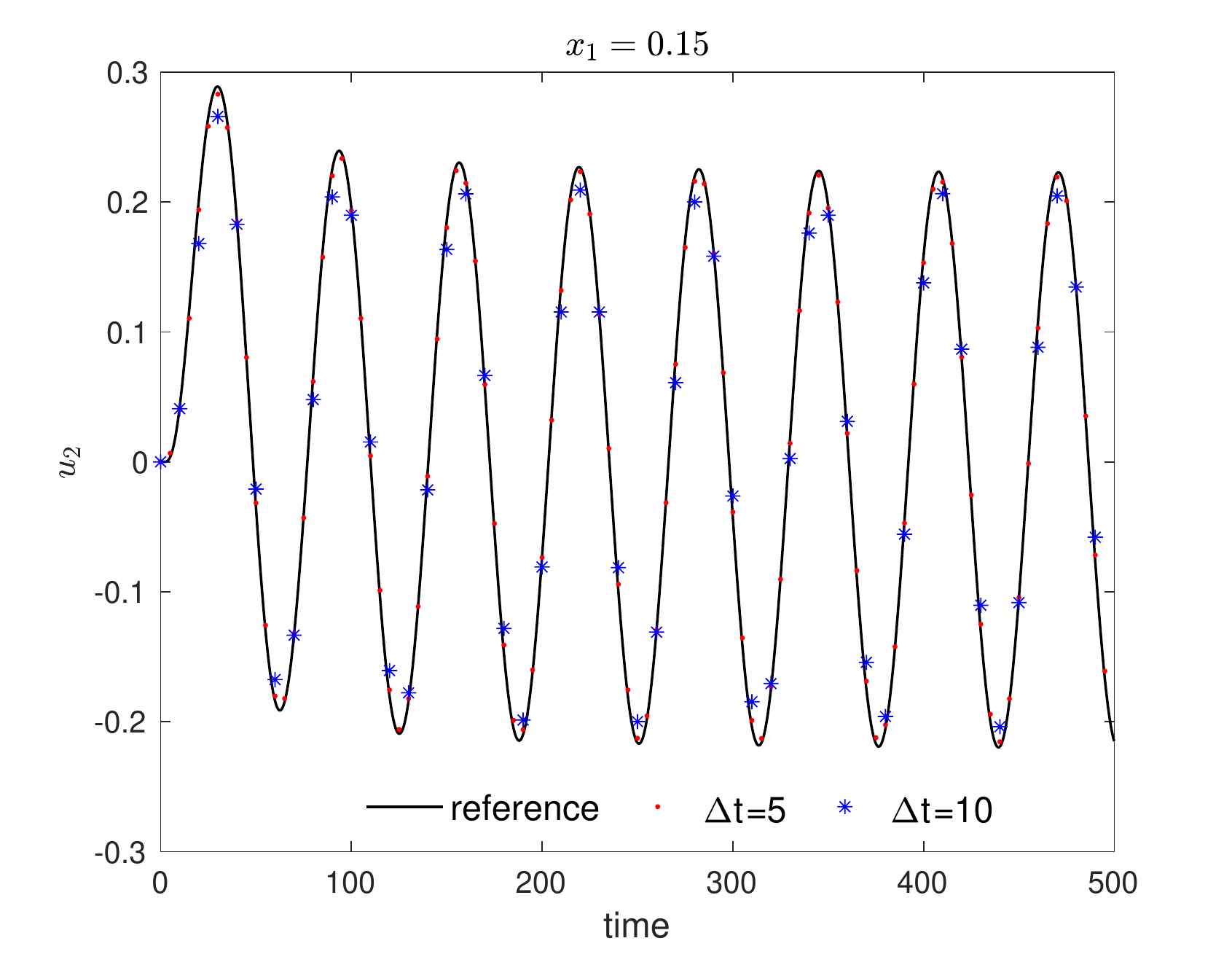}
	\includegraphics[scale=0.37]{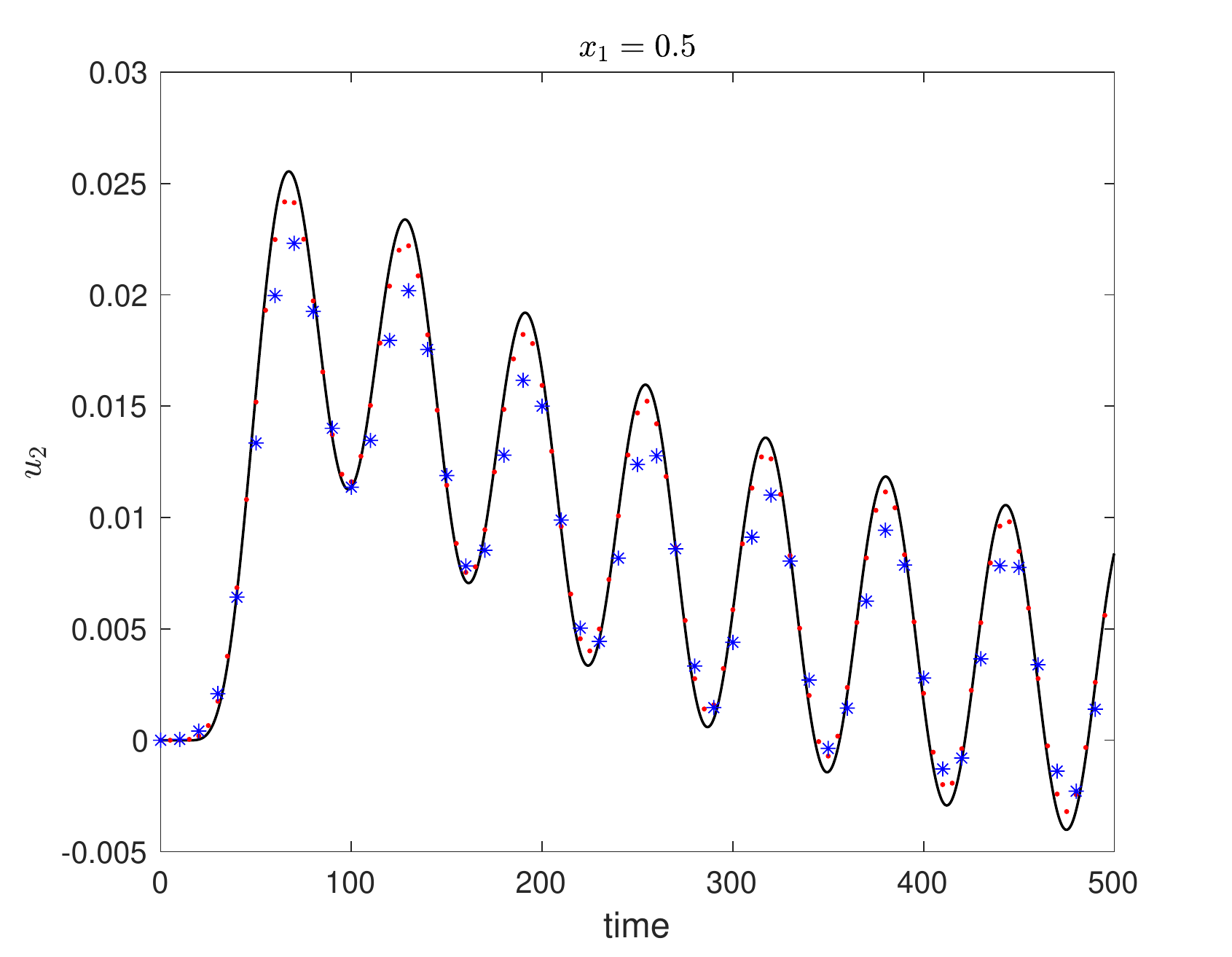}
	\caption{
		Velocity profiles in the oscillating Couette flow at $x_1=0.15$ and 0.5, when the BGK model with $\delta_{rp}=1000$ is solved by GSIS-I with $\Delta{t}=5$ and 10, respectively. The reference solution is obtained from GSIS-I with  $\Delta{t}=0.1$. Both kinetic and synthetic equations are solved by the Crank-Nicolson scheme, when the spatial region is discretized by Eq.~\eqref{spatial_discretization} with $N=121$.   
	}
	\label{fig:Couette_time}
\end{figure}

Finally, we assess the temporal accuracy of GSIS-I. It is seen from Fig.~\ref{fig:Couette_time} that even when the time step is much large than the mean collision time of gas molecules, say, one sixth of the oscillation frequency, the phase of the velocity is preserved after 10 oscillation periods. 

\section{Conclusions}\label{section_summary}

In summary, we have proposed two general synthetic iterative schemes to solve the gas kinetic equations efficiently and accurately. First, our rigorous Fourier stability analysis has shown that the GSIS enables fast convergence in the entire range of Knudsen number and time step. Second, the Chapman-Enskog expansion has been used to prove that, in the continuum flow regime, GSIS-I asymptotically preserves the Navier-Stokes-Fourier equations when the spatial and temporal step is $\Delta{t}, \Delta{x}\sim{O(1)}$, provided that the NSF equations can capture the hydrodynamics at this spatial grid size and temporal step. For GSIS-II, the spatial and temporal step should be made smaller to recover the NSF equations exactly, e.g., $\Delta{t}, \Delta{x}\sim{O(\sqrt{\text{Kn}})}$ when the kinetic equation is solved by a second-order accuracy temporal and spatial scheme.

From the analytical and numerical results we conclude that, in order to construct a fast converging and asymptotic preserving scheme for rarefied gas flows, it is necessary to couple the gas kinetic equation with the macroscopic synthetic equations. And in order to construct the GSIS-I scheme, it is necessary to reach at least the Grad 13 moment equations, where the highest order velocity moments are calculated directly from the numerical solution of kinetic equation. For other kinetic systems, say, for radiative heat transfer, phonon dynamics, we need the next-level moment system beyond the equations derived from the conservation laws. We plan to test this conjecture in other kinetic systems in the near future.


%


\bibliographystyle{siamplain}
\bibliography{thesisBib}

\end{document}

%% file: tmp_GSIS_HOLO_header.tex
	\title{General synthetic iterative scheme for unsteady rarefied gas flows
		 \thanks{Submitted to the editors DATE.
	}}
	
	\author{
		 Lei Wu\thanks{Department of Mechanics and Aerospace Engineering, Southern University of Science and Technology, Shenzhen 518055, China (\email{wul@sustech.edu.cn}) }
	}

	\headers{Unsteady GSIS: fast convergence and asymptotic  preserving}{Lei Wu}

%% file: tmp_GSIS_HOLO_abstract.tex
		\begin{abstract}
	In rarefied gas flows, the spatial grid size could vary by several orders of magnitude in  a single flow configuration (e.g., inside the Knudsen layer it is at the order of mean free path of gas molecules, while in the bulk region it is at a much larger hydrodynamic scale). Therefore, efficient implicit numerical method is urgently needed for time-dependent problems.
	However, the integro-differential nature of gas kinetic equations poses a grand challenge, as the gain part of the collision operator is non-invertible. Hence an iterative solver is required in each time step, which usually takes a lot of iterations in the (near) continuum flow regime where the Knudsen number is small; worse still, the solution does not asymptotically preserve the fluid dynamic limit when the spatial cell size is not refined enough.
Inspired by our general synthetic iteration scheme for steady-state solution of the Boltzmann equation, we propose two numerical schemes to push the multiscale simulation of unsteady rarefied gas flows to a new boundary, that is, the numerical solution not only converges within dozens of iterations in each time step, but also asymptotic preserves the Navier-Stokes-Fourier limit at coarse spatial grid, even when the time step is large (e.g., in the extreme slow decay of two-dimensional Taylor vortex, the time step is at the order of vortex decay time). The properties of fast convergence and asymptotic preserving of the proposed schemes are not only rigorously proven by the Fourier stability analysis, but also demonstrated by solid numerical examples.	
		\end{abstract}
		
		\begin{keywords}
			gas kinetic equation, fast convergence, asymptotic preserving
		\end{keywords}
		
		\begin{AMS}
			76P05, 
			65L04, 
			65M12 
			
		\end{AMS}